\documentclass[sigconf]{acmart}

\AtBeginDocument{
  \providecommand\BibTeX{{
    \normalfont B\kern-0.5em{\scshape i\kern-0.25em b}\kern-0.8em\TeX}}}

\copyrightyear{2026}
\acmYear{2026}
\setcopyright{cc}
\setcctype{by}
\acmConference[UIST '26]{The 39th Annual ACM Symposium on User Interface Software and Technology}{November 02--05, 2026}{Detroit, MI, USA}
\acmBooktitle{The 39th Annual ACM Symposium on User Interface Software and Technology (UIST '26), November 02--05, 2026, Detroit, MI, USA}
\acmDOI{10.1145/3830398.3830699}
\acmISBN{979-8-4007-2856-3/2026/11}
\usepackage{multirow}
\usepackage{booktabs,tabularx,array,makecell,siunitx,float}

\usepackage{color,soul}
\usepackage{booktabs} %much nicer tables
\usepackage{subfig} %for putting multiple subimages in one figure, with individual captions
\usepackage{graphicx}
\usepackage[utf8]{inputenc}
\usepackage[T1]{fontenc}
\usepackage{multirow}

\usepackage{siunitx}
\usepackage{enumitem}

\newcommand{\hide}[1]{}

\sethlcolor{yellow}
\ifdefined\scifihidecomments
    \newcommand{\cz}[1] {}
    \newcommand{\hyunc}[1] {} % Hyunchul
    \newcommand{\yax}[1] {} % Yaxuan
    \newcommand{\sony}[1] {} % Songyun
    \newcommand{\ReviewerFeedback}[1] {}  
    \newcommand{\fran}[1] {}
    \newcommand{\rz}[1]{}
    \newcommand{\lw}[1] {}
    \newcommand{\mose}[1] {} % Mose
    \newcommand{\ke}[1] {} % Ke
\else
    \definecolor{burntorange}{rgb}{0.8, 0.33, 0.0}
    \definecolor{cadmiumgreen}{rgb}{0.0, 0.42, 0.24}
    \definecolor{cobalt}{rgb}{0.0, 0.28, 0.67}
    \definecolor{amber}{rgb}{1.0, 0.75, 0.0}
    \definecolor{fashionfuchsia}{rgb}{0.96, 0.0, 0.63}
    \definecolor{brightcerulean}{rgb}{0.11, 0.67, 0.84}
    \definecolor{frenchblue}{rgb}{0.0, 0.45, 0.73}
    \definecolor{darkslateblue}{rgb}{0.28, 0.24, 0.55}
    \definecolor{cerulean}{rgb}{0.0, 0.48, 0.65}
    \definecolor{darkpastelgreen}{rgb}{0.01, 0.75, 0.24}
    \newcommand{\cz}[1] { \textcolor{red}{[\hl{cheng:} {#1}}]}
    \newcommand{\hyunc}[1] { \textcolor{burntorange}{[{\hl{hyunc:}} {#1}}]}
    \newcommand{\yax}[1] { \textcolor{magenta}{[{\hl{yaxuan:}} {#1}]}}
    
    \newcommand{\sony}[1] { \textcolor{blue}{[{\hl{songyun:}} {#1}]}}
    \newcommand{\ReviewerFeedback}[1] { \textcolor{brightcerulean}{[{Reviewer Feedback:} {#1}}]}
    \newcommand{\fran}[1]{\textcolor{burntorange}{[{francois:}{#1}}]}
    \newcommand{\rz}[1]{\textcolor{teal}{[{Ruidong: }{#1}]}}
    \newcommand{\lw}[1]{\textcolor{fashionfuchsia}{[{liuwei:}{#1}}]}
    \newcommand{\mose}[1]{\textcolor{burntorange}{[{mose:}{#1}}]}
    \newcommand{\ke}[1] { \textcolor{red!55!yellow}{[{Ke:} {#1}}]}

    \definecolor{CAT-comment}{rgb}{0.7, 0, 0}
    \definecolor{GL-comment}{rgb}{0.0, 0.54, 0.8}
    \definecolor{JW-comment}{rgb}{0.54, 0.54, 0.8}
    \definecolor{sensing-color}{rgb}{0.93, 0.45, 0.18}
    \definecolor{feedback-color}{rgb}{0.16, 0.39, 0.96}
    \definecolor{reconfig-color}{rgb}{1, 0.25, 0.96}
\newcommand{\thePlatform}{Sensorimotor Stickies}

\newcommand{\sensing}[1]{{\color{sensing-color} #1}}
\newcommand{\feedback}[1]{{\color{feedback-color} #1}}
\newcommand{\reconfig}[1]{{\color{reconfig-color} #1}}

\newcommand{\etal}{et al.~}
\fi

\ifdefined\scifiblind
    \newcommand{\blind}[1]{[omitted for blind review]}
\else
    \newcommand{\blind}[1]{#1} %for camera ready (not blinded)
\fi

\begin{document}
\title[\thePlatform{}]{\thePlatform: A Reconfigurable On-Body Platform for Closed-Loop Sensorimotor Training}

\author{Tianhong Catherine Yu}
\email{ty274@cornell.edu}
\affiliation{
  \institution{Cornell University}
  \city{Ithaca}
  \state{New York}
  \country{USA}
}

\author{Jiwei Zheng}
\email{jzheng26@uw.edu}
\affiliation{
  \institution{Univirsity of Washington}
  \city{Seattle}
  \state{Washington}
  \country{USA}
}

\author{Chi-Jung Lee}
\email{cl2358@cornell.edu}
\affiliation{
  \institution{Cornell University}
  \city{Ithaca}
  \state{New York}
  \country{USA}
}

\author{Qifeng Yang}
\email{qyang24@uw.edu}
\affiliation{
  \institution{Univirsity of Washington}
  \city{Seattle}
  \state{Washington}
  \country{USA}
}

\author{Tingyu Cheng}
\email{tcheng2@nd.edu}
\affiliation{
  \institution{University of Notre Dame}
  \city{Notre Dame}
  \state{Indiana}
  \country{USA}
}

\author{Qiuyue (Shirley) Xue}
\email{qiuyue@purdue.edu}
\affiliation{
  \institution{Purdue Univeristy}
  \city{West Lafayette}
  \state{Indiana}
  \country{USA}
}

\author{Cheng Zhang}
\email{chengzhang@cornell.edu}
\affiliation{
  \institution{Cornell University}
  \city{Ithaca}
  \state{New York}
  \country{USA}
}

\author{Yiyue Luo}
\email{yiyueluo@uw.edu}
\affiliation{
  \institution{Univirsity of Washington}
  \city{Seattle}
  \state{Washington}
  \country{USA}
}

\settopmatter{authorsperrow=4}
\renewcommand{\shortauthors}{Yu, et al.}

\begin{abstract}
Closed-loop sensorimotor training systems can improve learning by sensing movement and delivering real-time feedback; yet, most existing systems are still rebuilt as task-specific one-offs, even though much of the wearable sensing-and-feedback stack remains the same.
We present Sensorimotor Stickies, a reconfigurable on-body platform that treats inertial \& tactile sensing and vibrotactile feedback as modular stickies that can be patched onto the body as needed. 
The platform combines miniaturized adhesive sensing and feedback modules, reusable low-power firmware \& BLE infrastructure that support raw streaming and motor control without task-specific firmware rewrites, and a body-centered configuration model realized through a companion mobile app for placement, calibration, feedback authoring, and closed-loop training use. 
Together, these components enable reconfiguring closed-loop feedback across tasks, users, and feedback setups using the same platform. 
We evaluate the platform through: configured task examples showing breadth across training scenarios; a practitioner elicitation and co-configuration study (N=6) grounding the platform in meaningful real-world uses while probing fit and limits; an end-user study (N=10) showing first-time setup, calibration, and within-task feedback configurations for a preconfigured task; and technical characterization showing feasibility for untethered closed-loop deployment.
\end{abstract}




\begin{teaserfigure}
\vspace{-12pt}
  \includegraphics[width=\textwidth]{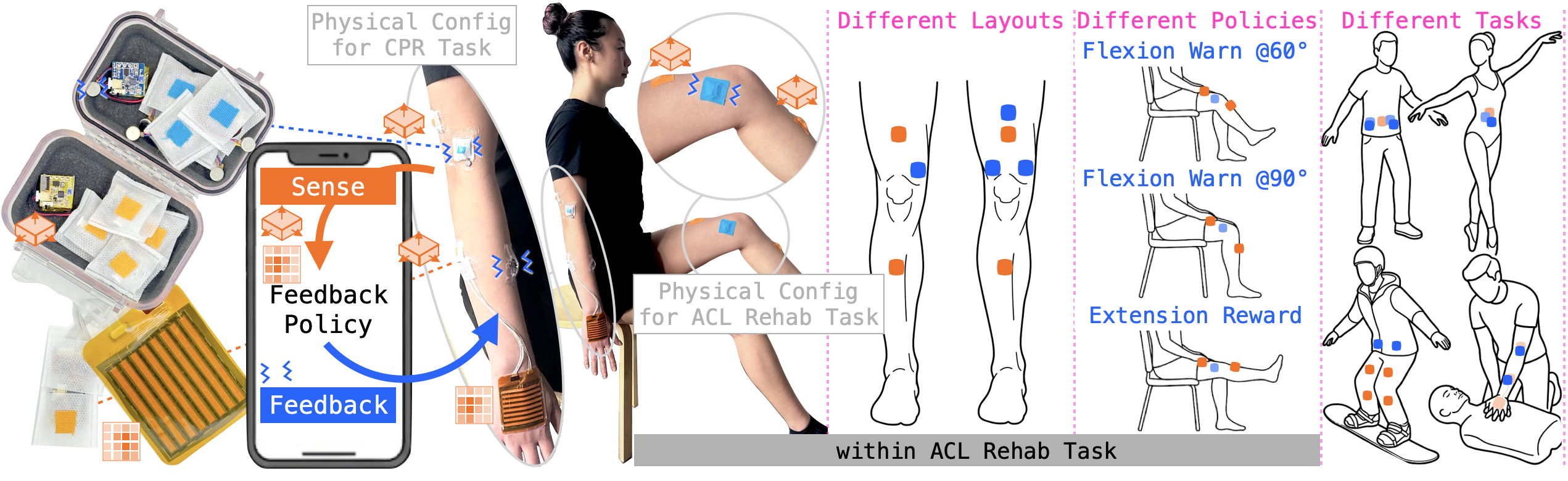}
  \vspace{-20pt}
  \caption{The \thePlatform{} platform enables \reconfig{reconfigurable} closed-loop on-body \sensing{inertial and (optional) tactile sensing} and \feedback{vibrotactile feedback} for sensorimotor training. The platform combines \sensing{sensing} and \feedback{feedback} stickies that can be taped to skin, with a mobile app, illustrated here with a CPR compression task (on the arm with 4 stickies) and an Anterior Cruciate Ligament (ACL) post-surgical rehab task (on the leg with 3 stickies). Taking the ACL rehab task as an example, the same underlying hardware, firmware, and software infrastructure can be \reconfig{reconfigured across layouts, policies, and training tasks}.}
  \label{fig:teaser}
\end{teaserfigure}

\maketitle
\vspace{-12pt}
\section{Introduction}
Sensorimotor training relies on iterative perception-action loops: learners move, receive feedback, adjust, and repeat. 
Dancers watch mirrors, patients respond to a therapist’s correction, and athletes adapt based on a coach’s cue. 
When feedback is delivered during movement, learners can respond immediately rather than only after the fact. A growing body of work has therefore explored closed-loop systems that sense task-relevant body-state and provide feedback in real time during practice across rehabilitation~\cite{review-real-time-biomechanical-feedback}, skill acquisition~\cite {AdaptiveEMS, Hapticus}, ergonomics~\cite{sitting-posture-review}, and procedural training~\cite{CPR-glove}.

Among these, a deliberately constrained but practically important class centers on on-body systems for sensorimotor training, relying on motion sensing and, in some cases, physical contact sensing.
Inertial Measurement Units (IMUs) capture body segment motion and joint relationships, while tactile sensing extends the stack when physical interaction matters.
Vibrotactile feedback provides timely, untethered on-body cues during movement, avoids directly competing for visual/auditory attention, and can be deployed with relatively compact hardware.
Using this stack, researchers have seen promising results in rehabilitation and motor relearning~\cite{review-real-time-biomechanical-feedback, vibrotactile-rehab-review}, technique training (e.g., sports~\cite{vibrotactile-sports-review} and dance~\cite{GuiDance}), occupational ergonomics~\cite{inertial-work-activities-review}, and other structured training tasks~\cite{CPR-band}.

However, despite repeated demonstrations of value, these systems are most often built as fixed implementations: each tailored to one task, one sensing algorithm, one feedback logic, and one deployment context.
Yet sensorimotor training inherently demands adaptation~\cite{flexible-wearables-physical-training,agumented-feedback-motor-learning}.
Different tasks require different sensed variables, placements, and feedback logic.
Even within a single task, effective feedback must evolve as the learner progresses~\cite{attentional-focus, frequent-feedback-degrade-learning} and depends on individual perception that only becomes clear during use~\cite{kinematic-vibrotactile-mapping}.
Across training scenarios, the core technology remains similar, but the configuration changes. 
What is therefore missing is a way to reconfigure the same underlying sensing-and-feedback stack as tasks, users, and training goals change.

\begin{figure}[t]
\includegraphics[width=\columnwidth]{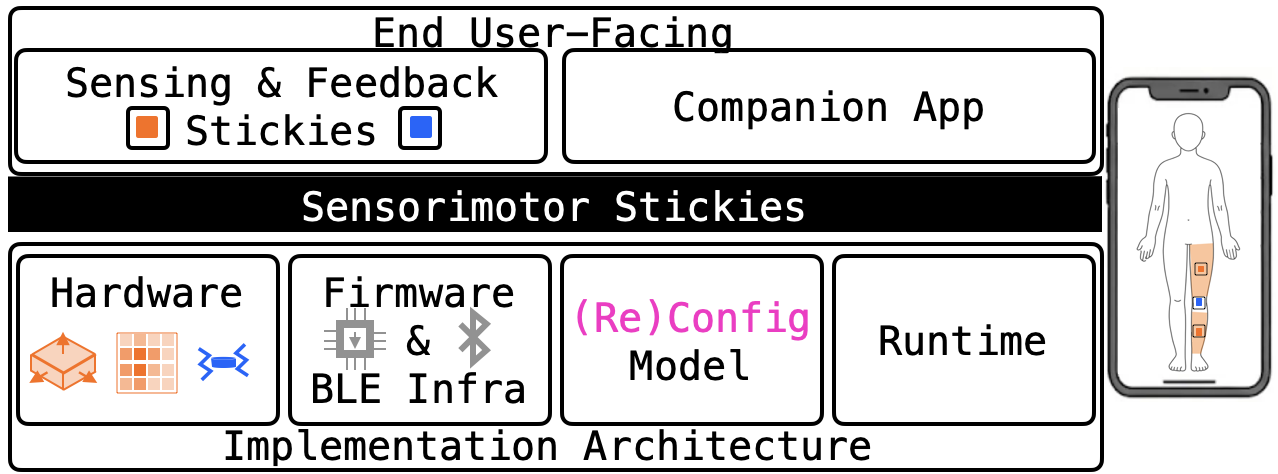}
  \caption{\thePlatform{} Overview.}

  \label{fig:platform-overview}
\vspace{-20pt}
\end{figure}

Thus, our goal is not modality breadth, but reconfiguration of sensing and sparse on-body feedback.
Inspired by the on-demand, self-applicable, and adaptable nature of adhesive bandages, we \textbf{envision sensorimotor training augmentation as something one can patch onto the body as needed}. 
To this end, we present \textbf{Sensorimotor Stickies, a reconfigurable platform for closed-loop on-body vibrotactile feedback in sensorimotor training}, as shown in Fig.~\ref{fig:teaser}.
The platform combines 
(1) a physical wearable layer of miniaturized, low-power adhesive sensing and feedback modules designed for unobtrusive wear during active movement, 
(2) a reusable embedded layer with Bluetooth Low Energy (BLE) infrastructure for raw sensor streaming and motor control across tasks without firmware rewrites, and 
(3) a body-centered configuration model and runtime for setup, calibration, feedback authoring, and closed-loop execution during use. 
As shown in Fig.~\ref{fig:platform-overview}, end users encounter this platform through sensing and feedback stickies together with a companion mobile app, which serves as the user-facing realization of the configuration model and runtime.
Together, these components turn a repeatedly useful but typically one-off class of wearable systems into a reconfigurable platform.
\vspace{-2pt}
Guided by Ledo~\etal{}~\cite{toolkit-evaluation}, our multi-strategy evaluation includes four parts:
\textbf{(1) reconfiguration breadth by presenting task configurations} based on prior work and instantiated through our app workflow;
\textbf{(2) grounded applicability through a practitioner elicitation and co-configuration study} with movement-domain practitioners, showing that the platform can be mapped to meaningful scenarios from their own practice while probing opportunities and limitations;
\textbf{(3) first-time setup and within-task reconfiguration, through an end-user study}, suggesting that users can generally attach the stickies, complete calibration, and personalize feedback placement and vibration patterns for a preconfigured task;
and \textbf{(4) technical feasibility by characterization} of BLE communication latency, packet loss, and energy use, finding mean BLE latency of 38.2 ms with the phone on the body and 42.5 ms at 5 m, negligible packet loss, and battery behavior consistent with full-day continuous use and multi-day sparse, event-driven use.
Each part addresses a distinct concern: whether the platform can express diverse tasks, whether practitioners would use it for real goals, whether end users can operate it, and whether the system performs well enough for untethered deployment.

In summary, the main contributions of this paper are:
\begin{itemize}[leftmargin=*, nosep]
    \item \textbf{A practical wearable infrastructure for closed-loop on-body feedback} consisting of miniaturized adhesive sensing and feedback modules, together with reusable low-power firmware and BLE infrastructure designed to persist beyond any single front-end, supported by technical characterization.
    \item \textbf{A reconfigurable platform for sensorimotor training} that supports reconfiguration through a body-centered configuration model and runtime, realized for end users through a companion mobile app, and demonstrated through task configurations spanning prior work replication and practitioner-grounded scenarios.
    \item \textbf{Empirical evidence from a practitioner study and an end-user study} showing how movement practitioners map grounded training scenarios onto the platform’s configuration model, and demonstrating the feasibility of first-time setup, calibration, and within-task feedback reconfiguration for a preconfigured task.
\end{itemize}

The platform is open-sourced\footnote{\url{https://stickies.catherineyu.com}}, including manufacturable hardware, firmware, and app resources to ease reproducing and building on this class of real-time feedback systems.

\section{Related Work}
\thePlatform{} targets reconfigurable closed-loop on-body sensing and vibrotactile feedback for sensorimotor training tasks.
Prior research has explored many combinations of sensing (e.g., marker-based motion capture~\cite{Ariadne, TIKL}) and feedback (e.g., audio, visual, or haptic~\cite{Sensorimotor-Learn-VR, AdaptiveEMS}).
Among them, our focused class is portable, unobtrusive, and can provide cues without directly competing for visual or auditory attention~\cite{Hapticus,agumented-feedback-motor-learning}.
We therefore focus this section on prior work most closely related to \thePlatform{}: 
(1) closed-loop vibrotactile feedback systems for sensorimotor training, 
and (2) tools for building wearable sensing and feedback systems.
\\

\begin{table}[t]
\caption{Closed-loop wearable sensing and vibrotactile feedback systems across sensorimotor training domains.}
\label{tab:related-work-domains}
\centering
\small
\setlength{\tabcolsep}{4pt}
\renewcommand{\arraystretch}{1.3} % Added for better spacing between dense rows
\begin{tabularx}{\columnwidth}{@{}>{\raggedright\arraybackslash}p{0.3\columnwidth} X@{}}
\toprule
\textbf{Domain} & \textbf{Feedback roles \& representative tasks} \\
\midrule
\textbf{Rehabilitation, Motor Relearning}\newline (Reviews~\cite{review-real-time-biomechanical-feedback, vibrotactile-rehab-review})
& Cues deviation or reinforces desired movement for balance~\cite{vibrotactile-for-standing}, gait retraining~\cite{Ariadne}, and upper-limb guidance~\cite{TIKL, vibrotactile-arm-motions-lessons}. \\
\midrule

\textbf{Technique Training}\newline (Reviews~\cite{vibrotactile-sports-review})
& In-motion cues for refining technique without interrupting practice for sports~\cite{VibroBits, tactile-for-snowboard}, fitness~\cite{GymSoles, yoga-vibrotactile}, dance~\cite{GuiDance}, and music~\cite{Musicjacket, vibrotactile-deign-music}. \\
\midrule

\textbf{Occupational, Ergonomics}\newline (Reviews~\cite{inertial-work-activities-review, wearable-occupational-review, sitting-posture-review})
& Cues when posture or movement enters an undesirable range as repetitive-work habits~\cite{imu-vibrotacrtile-ergonomics-intervention} while sitting~\cite{seated-vibrotactile-garment} and lifting~\cite{vibrotacitle-ergonomic}. \\
\midrule

\textbf{Procedural Skill Training}
& Cues deviation from required form or timing during structured practice such as CPR compression~\cite{CPR-glove, CPR-band} and needle handling~\cite{needle-wristband}. \\
\bottomrule
\end{tabularx}
\vspace{-10pt}
\end{table}

\noindent\textbf{Closed-Loop Vibrotactile Feedback Systems.}
\\
\noindent Although vibrotactile feedback is not always the most directive or effective modality for every training objective, prior work shows that it offers a distinctive combination of portability, comfort, and user agency, making it a valuable option for wearable real-time cueing, as detailed in Table~\ref{tab:related-work-domains}.
Since Lieberman~\etal{}showed that TIKL, a motion-capture-driven vibrotactile feedback system, improved arm-motion learning~\cite{TIKL}, similar systems have been studied across rehabilitation~\cite{vibrotactile-for-standing, dots}, sports training~\cite{tactile-for-snowboard, VibroBits}, ergonomics~\cite{vibrotacitle-ergonomic, sitting-posture-review}, and skill acquisition~\cite{Musicjacket, vibrotactile-arm-motions-lessons}.
Across these systems, the underlying technical pattern is similar: capturing body-state, deriving task-relevant sensing metrics, and delivering sparse on-body vibrotactile cues, yet each application requires application-specific choices in sensing configuration, feedback layout, and control logic.
This repeated pattern suggests that the key challenge is not only reusable hardware, but reconfiguring the sensing-feedback loop itself.
Recognizing the need to support different rehabilitation movements, Xu~\etal{}developed Dots, reusable wireless nodes that can be configured for inertial sensing and/or vibrotactile feedback~\cite{dots}. 
Dots demonstrated that the same hardware substrate could support both balance (when placed around the waist) and gait (when placed on the foot) training.
\thePlatform{} extends this reconfigurable hardware direction beyond predefined tasks by making the sensing-feedback loop reconfigurable too.
Through a shared body-centered model, for a task, users choose sensing metrics such as joint angles and tactile presence, calibrate them for a task, and map them to personalized feedback locations and policies without task-specific engineering.
\\

\noindent\textbf{Tool Support for Configurable On-Body Feedback.}

\noindent A separate line of work has lowered the barrier to building wearable sensing and feedback systems, but typically supports only parts of the closed-loop pipeline targeted by~\thePlatform{}.
For sensorimotor training, these parts must be connected into a real-time loop that non-technical users can configure and use.

\underline{General-purpose wearable prototyping toolkits}
have reimagined prototyping on-body interactive systems. 
From early LilyPad~\cite{LilyPad} for e-textile prototyping and miniaturized commercial platforms (e.g., XIAO and QT Py) to recent on-skin interface construction toolkits~\cite{Skinlink}, these toolkits lower the barrier for creative exploration. 
However, they are designed for expressive breadth, leaving programming decisions to the designer or developer for each new use case.
\thePlatform{} takes the opposite approach: by targeting closed-loop sensorimotor feedback, we make hardware and firmware choices (e.g., miniaturized adhesive form factor, low-power raw streaming without on-device processing) that would not be appropriate for a general-purpose toolkit but are well suited to unobtrusive, long-term uses across configurations~\cite{Thermal-earring}.

\underline{Sensing and motion interpretation tools} focus on capturing or interpreting body states.
Commercial MoCap platforms~\cite{XSens, OptiTrack, Vicon} and biomechanics software (e.g., OpenSim~\cite{OpenSim}) support detailed kinematic and musculoskeletal modeling, while recent open-source platforms~\cite {WiReSens} lower the barrier to building wireless tactile sensing.
These tools are powerful for measuring and interpreting body-state, but they often require specialized expertise, expensive or complex infrastructure, and substantial engineering to integrate into real-time feedback systems, making them less accessible to non-technical users seeking configurable training support.

\underline{Vibrotactile haptic rendering toolkits} support the creation, prototyping, or sharing of vibrotactile patterns across devices and application contexts, making expressive tactile rendering accessible~\cite{vibraforge, VHP, Tape-Tics, TactJam, Sound2Haptic, Scene2Hap}.
Similar to the sensing tools, the vibrotactile toolkits alone can be building blocks within larger feedback systems with careful engineering.
In contrast, \thePlatform{} focuses on closed-loop training scenarios where feedback is typically sparse and event-driven for effective interpretation~\cite{CPR-glove, kinematic-vibrotactile-mapping}, making configuration of sensing, placement, and triggering more central than rich haptic composition.

\underline{Rule-based authoring tools}
enable non-programmers to specify behavior through conditions and corresponding responses.
The trigger-action programming paradigm is particularly effective in authoring IoT interactions~\cite{trigger-action-programming, customize-domain-specific-tools, aveni2025generative}. 
SmartFit Rule Editor~\cite{smartfit-rule-editor} brings this paradigm to the wearable fitness domain and enables coaches and trainers to define rules over wearable data streams. 
However, rule authoring alone does not solve the closed-loop problem: placing sensors on the body, calibrating sensed metrics, choosing feedback locations, and executing the loop on wearable hardware.
\thePlatform{} therefore extends beyond rule specification to an integrated wearable stack: miniaturized on-body sensing and feedback hardware, reusable runtime infrastructure, and a mobile workflow for setup, calibration, logic adjustment, and use.

\section{\thePlatform{}}
\thePlatform{} treats a practically important class of closed-loop wearable systems not as one-off builds, but as configurations of a shared platform.
Guided by the design goals (Sec.~\ref{sec:design-goals}), we describe the platform’s body-centered configuration model and how it supports reconfiguration across sensing metrics, feedback logic, and feedback layout in Sec.~\ref{sec:platform-overview}.
Sec.~\ref{sec:walkthrough} presents an ACL post-surgical rehabilitation walkthrough illustrating how the platform supports reconfigurations.

\subsection{Design Goals}\label{sec:design-goals}
\noindent\textbf{DG1. Support practical on-body use.}
The wearables should be practical to wear, attach, and calibrate.

\noindent\textit{DG1.1. Enable unobtrusive wear during movement.}
The stickies should be small, untethered, low-power, and unobtrusive enough for prolonged use during movement practice.

\noindent\textit{DG1.2. Make setup, placement, and calibration accessible.}
Users should be able to attach the devices and complete calibration despite placement variation and other messiness of actual use.

\noindent\textbf{DG2. Support (re)configuration for sensorimotor training.}
The platform should accommodate different tasks, users, and progression stages without rebuilding a new system each time~\cite{sports-wearables-design-space, flexible-wearables-physical-training}.

\noindent\textit{DG2.1. Reuse shared infrastructure.}
The same hardware and firmware should support different sensing, feedback, and policy needs without task-specific rewrites.

\noindent\textit{DG2.2. Author configurations in body- and task-level terms.}
Users should be able to specify body parts, sensed metrics, feedback locations, and feedback policies in high-level terms rather than low-level technical parameters~\cite{Intercorporeal-biofeedback}.
\begin{figure}[b]
\includegraphics[width=\columnwidth]{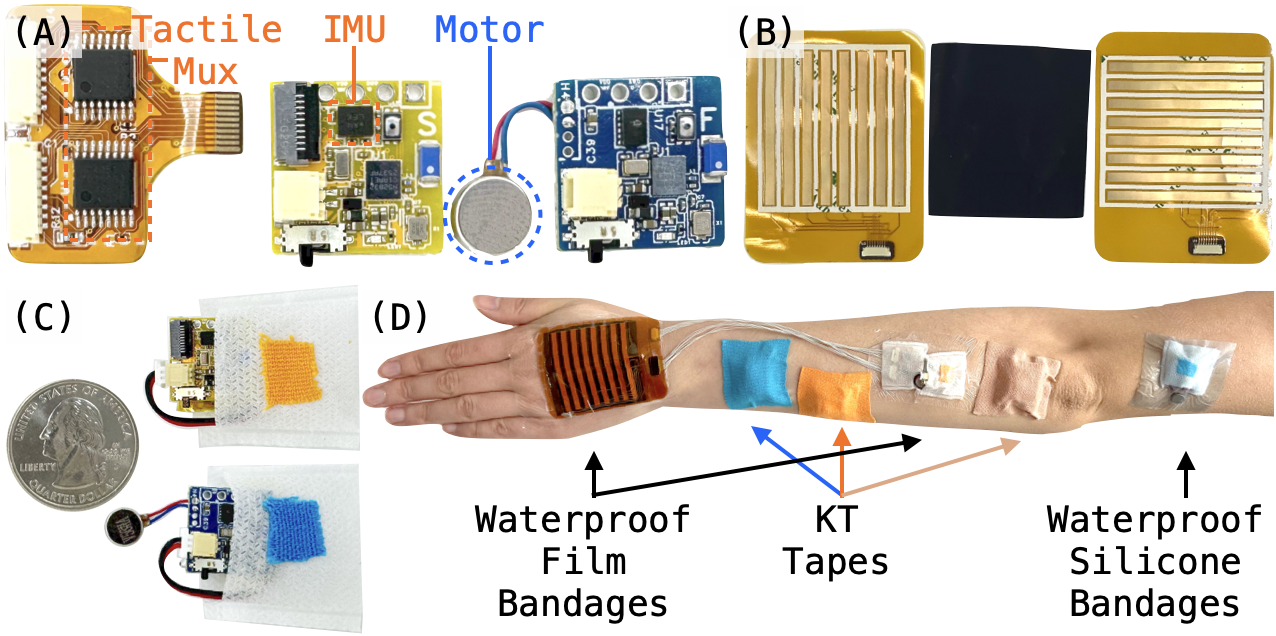}
\vspace{-15pt}
  \caption{(A) Sensing and feedback PCBs and the optional FPCB tactile sensing expansion circuit. (B) Tactile sensing patch, made of FPCB and a middle Velostat layer. (C) Stickies in pouches. (D) Adhering with different skin-safe tapes.}
  \label{fig:wearable-stickies}
\end{figure}

\subsection{Platform Overview \& Realization}\label{sec:platform-overview}
\thePlatform{} combines adhesive sensing and feedback stickies, reusable embedded infrastructure, and a body-centered configuration model realized through a companion mobile app.
See Appendix~\ref{sec:appendix-implementation} for additional implementation details.

The \textbf{wearable layer}, as shown in Fig.~\ref{fig:wearable-stickies}, consists of two small BLE stickies: sensing stickies and feedback stickies.
Sensing stickies integrate a 6-axis IMU and an expansion port for optional add-on sensing like the thin tactile sensing-matrix patches.
Feedback stickies integrate a motor driver and up to two vibrotactile motors.
All stickies are independent BLE modules that can be placed where needed on the body.
Each sticky measures approximately 2.5 × 2.5 × 0.8 cm, weighs about 2.5g, including a detachable, rechargeable LiPo battery (70 mAh, 2g).
The PCBs (See Tables~\ref{tab:sensing_components_cost} and \ref{tab:feedback_components_cost} and Fig.~\ref{fig:pcb-details} for details) are enclosed in breathable, water-resistant pouches and adhered using skin-safe tapes such as medical, kinesiology, or body tapes, supporting practical recharging, replacement, and unobtrusive wear during movement.

\begin{figure}[t]
\includegraphics[width=\columnwidth]{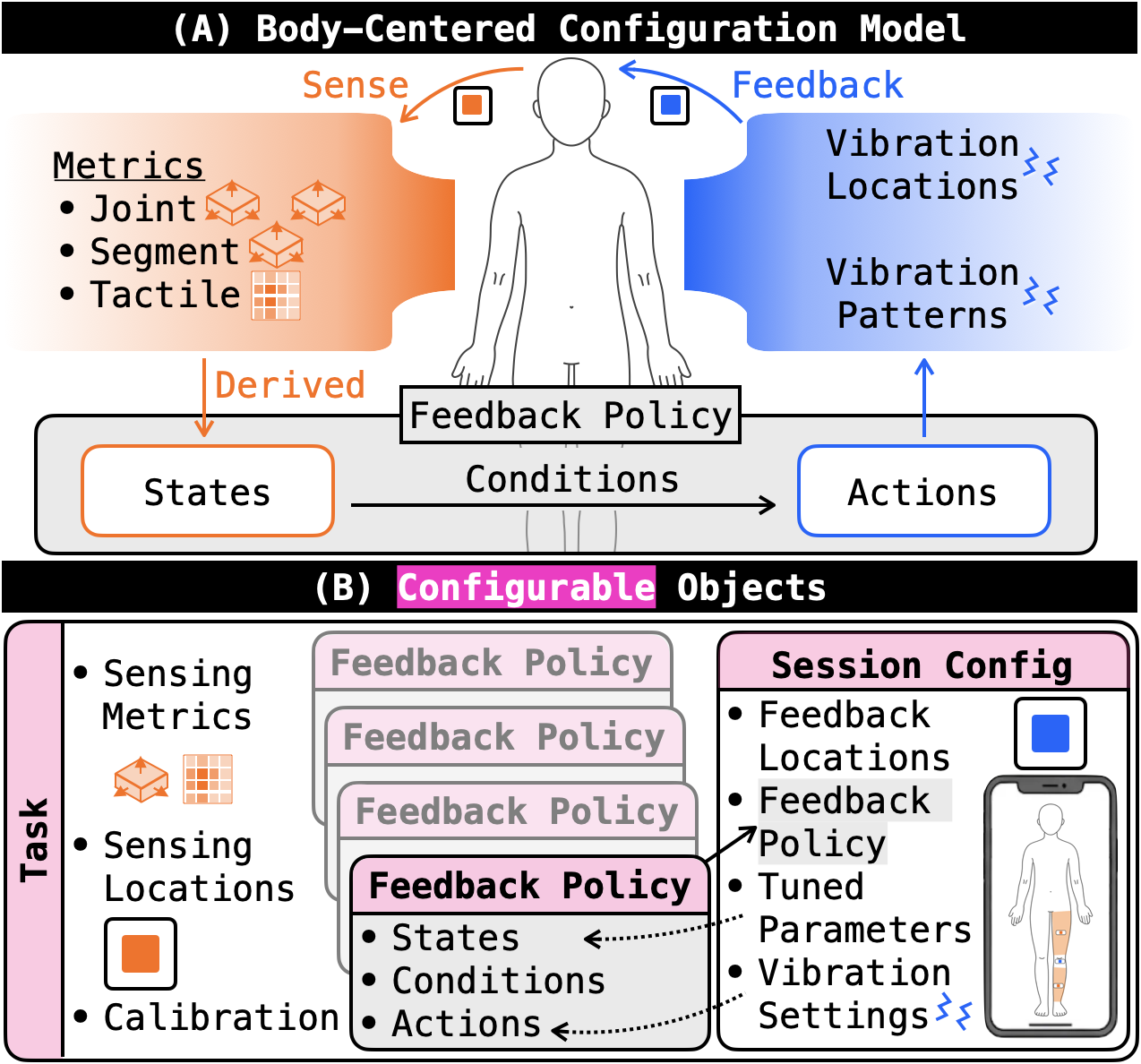}
\vspace{-15pt}
  \caption{\thePlatform{}’ (A) configuration model and (B) configurable objects.}
  \vspace{-15pt}
  \label{fig:configuration-model}
\vspace{-10pt}
\end{figure}
\begin{figure*}[t]
\includegraphics[width=\textwidth]{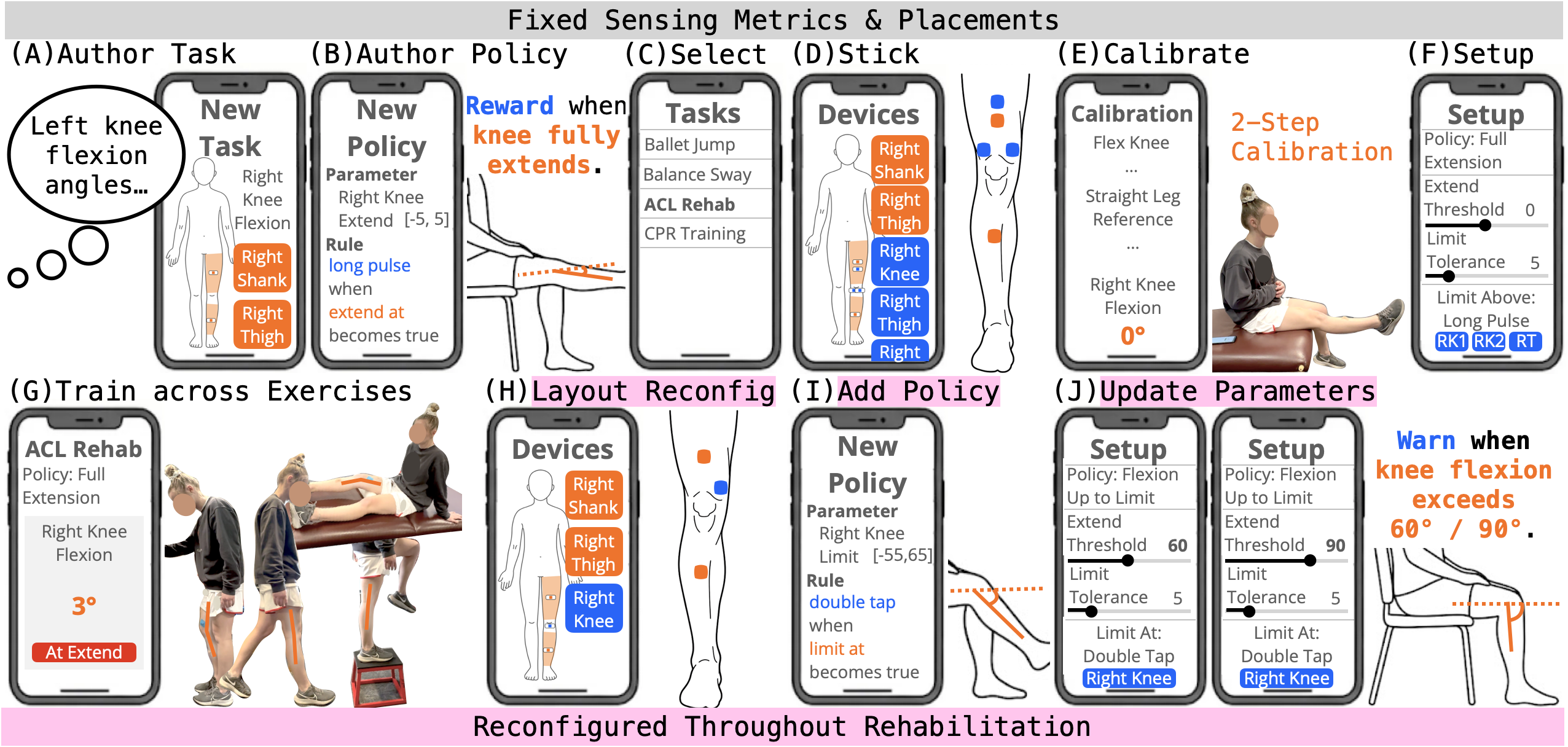}
\vspace{-15pt}
\caption{Using scenario walkthrough for ACL rehab.  
The UIs are modified based on the app design (see Fig.~\ref{fig:ui_screenshots}) for readability.}
\vspace{-5pt}
  \label{fig:walkthrough}
\end{figure*}

The \textbf{embedded layer} uses reusable role-specific firmware on the same BLE platform, detailed in Table~\ref{tab:ble_gatt}, with separate sensing and feedback roles.
BLE services expose the sticky type and capabilities so the app can distinguish sticky types without manual device-type selection.
On boot and connect, sensing stickies automatically detect whether a tactile extension is attached.
Sensing firmware streams raw IMU data, and tactile data when present, while feedback firmware receives motor commands.
Together, this infrastructure supports reuse across tasks without task-specific firmware rewrites.

The platform is organized around a \textbf{body-centered configuration model}, as shown in Fig.~\ref{fig:configuration-model}(A), that describes sensing and feedback in anatomical terms rather than raw stickies IDs or data. 
Our configuration model separates the following components.
\begin{itemize}[leftmargin=*, nosep]
\item \textbf{Body-state metrics} are task-relevant quantities computed from sensing stickies to provide interpretable movement or contact variables. We currently support 3 metric families: (1) \textbf{joint metrics} are computed from IMUs on 2 adjacent body segments and include joint angles (e.g., flexion and pronation) and angular velocities; (2) \textbf{segment metrics} are computed from 1 IMU and include orientation-(e.g., pitch and roll) and motion-derived (e.g., angular velocity and linear acceleration) metrics; and (3) \textbf{tactile metrics} are computed from the tactile matrix patch and include contact presence, total \& peak tactile values, contact area, and x \& y coordinates of the center of pressure. These metrics provide a constrained but reusable library of sensing abstractions. 

\item\textbf{Metric-specific calibration} translates raw sensor signals into the user-facing body-state metrics and enables practical non-expert placement variations (e.g., differences in sticky orientation and exact placement). For joint metrics, users perform a short calibration movement and a user-specified reference pose (e.g., straight arm): the system estimates functional joint axes, heading offsets, and body-relative frames (from the reference pose) so users do not need to precisely align IMU axes with anatomical axes during placement. The current implementation supports 1-DoF and 2-DoF joint-angle pipelines ~\cite{vqf, 1dof-algo, 2dof-algo}; for 3-DoF joints such as shoulders or hips, the system currently supports constrained task movements by treating the relevant motion as a 1-DoF joint task. For segment metrics, calibration uses a user-specified reference pose to express IMU orientation relative to the body-centered frame. For tactile metrics, calibration records a no-contact baseline and patch orientation so raw tactile readings can be converted into contact and spatial pressure summaries.

\item \textbf{Feedback locations \& patterns} configure the haptics experience. Users choose the feedback locations based on their preferences and assign the attached body parts when connecting the feedback stickies. Vibration patterns specify how a triggered action is rendered: vibration duration and vibration repetitions can be additionally customized (e.g., a long pulse and triple taps).
\end{itemize}

\noindent As illustrated in Fig.~\ref{fig:configuration-model}(B), these components are organized into 3 configurable objects:
\begin{itemize}[leftmargin=*, nosep]
\item A \textbf{Task} specifies which sensing stickies are required and where they are placed, what calibration the user performs, and which body-state metrics are computed. 
For example, a joint-angle task specifies the two body segments (e.g., thigh and shank for the knee) to instrument, the joint calibration (e.g., bending the knees for 10s and capturing the straight leg as the 0 \textdegree{} reference) to perform, and the joint metrics (e.g., knee flexion angle and knee flexion velocity) exposed to feedback policies.
Each task can have multiple Feedback Policies. 
\item A \textbf{feedback policy} defines a rule structure over the metrics provided by a task. In the current implementation, feedback policies are threshold-based rules. A policy defines named states over sensing metrics (e.g., \texttt{at\_target} when knee flexion is within a target range, or \texttt{hand\_centered} when tactile center-of-pressure coordinates fall within a target region), conditions (e.g., \texttt{when becomes true}) over those states (e.g., \texttt{knee\_flexion\_at\_target}), and actions that specify which vibration pattern should be triggered when the conditions are met. States can also be composed with logical operators to form compound states or conditions, allowing policies to express cases such as \texttt{at\_target AND moving} (as shown in Fig.~\ref{fig:edit-screenshots}) or tactile-region rules that combine x- and y-coordinate states (as shown in Fig.~\ref{fig:tactile-screenshots}). This rule-based abstraction is deliberately simple: many structured sensorimotor training tasks depend on monitoring a small number of movement or contact variables and cueing when they enter, are at, or leave an acceptable range. The current policy model does not support sliding-window aggregation or learned policies such as pose classification.

\item A \textbf{session configuration} instantiates a task with a selected policy for the session to specify feedback locations, tuned parameters, and vibration settings.
These session-level settings do not define a new policy from scratch; instead, they tune the selected policy by adjusting parameterized states (as shown in Fig.~\ref{fig:walkthrough}) and customizing how actions are rendered on the body.
\end{itemize}
At runtime (i.e., while training), the configured task, policy, and session settings are executed over live sensor streams: the system computes metrics and states, evaluates policy conditions, and sends vibrotactile commands to the selected feedback stickies.
The companion app exposes this model through a workflow for selecting tasks, connecting and assigning devices, calibrating sensing, choosing among a task's policies, tuning session settings, and running training sessions.

\vspace{-10pt}
\subsection{Using Scenario Walkthrough}\label{sec:walkthrough}
To concretize the platform, we present a scenario walkthrough, illustrated in Fig.~\ref{fig:walkthrough}, as 1 task with 2 policies, grounded in 2 tasks configured by athletic trainers in our study (detailed in Sec.~\ref{sec:practioner-study}).

Kate, a collegiate athlete, is undergoing post-surgical ACL rehabilitation with Alice, her athletic trainer (AT).
Sensorimotor Stickies do not replace Alice’s role.
Rather, the platform extends her cues into repeated movements that she cannot continuously supervise.

\textbf{Full Extension Reward}    (Fig.~\ref{fig:walkthrough}(A-G)). 
One early rehabilitation goal is to restore and eventually maintain full extension.
Alice configures a knee-angle task with a reward policy that vibrates when full extension is reached.
Kate places sensing stickies on her shank and thigh, and 3 feedback stickies around the knee and thigh because her leg sensation is still affected by surgery.
Kate performs 10 seconds of knee flexion for the knee tracking algorithm to identify the joint axis, then fully extends the knee to set the 0° reference.
Before starting, she tunes the policy by setting the extension threshold (e.g., within 5\textdegree{} of full extension) and vibration pattern.
She then performs full-extension exercises during sitting, standing, walking, and step-ups.
After training, she removes the stickies, discards the tapes, and reuses the electronics.

\textbf{Layout Reconfiguration} (Fig.~\ref{fig:walkthrough}(H)).
Over time, 3 feedback stickies may become unnecessarily strong, so Alice instead suggests using a single feedback sticky just above the inside of the knee near the vastus medialis, a muscle often emphasized during terminal extension, serving as biofeedback for full extension.

\textbf{Flexion Limit Warning} (Fig.~\ref{fig:walkthrough}(I-J)). 
Another important rehabilitation consideration is phased flexion limits. 
Using the same knee-angle task, Alice adds another policy that warns when Kate approaches the allowed flexion limit (e.g., 90\textdegree{}) during rehabilitation activities such as walking.
As Kate progresses, Alice updates this flexion limit with input from Kate’s physical therapist and surgeon.

\textbf{Sensorimotor Stickies for Another Athlete}. Because ACL rehab often follows standard protocols, similar policies can be reused for other athletes undergoing similar rehab, while still being personalized to each athlete’s feedback needs and progression.

This walkthrough illustrates policy and layout reconfiguration for a single metric in one rehabilitation context. 
Sec.~\ref{sec:applications} shows breadth across task configurations.
Sec.~\ref{sec:practioner-study} examines grounded practitioner uses.
Sec.~\ref{sec:end-user-study-results} studies end-user setup and reconfiguration.
Finally, Sec.~\ref{sec:technical-characterization} characterizes technical performance.

\section{Task Configurations based on Prior Work}\label{sec:applications}
\begin{figure}[b]
  \vspace{-12pt}
  \includegraphics[width=\columnwidth]
  {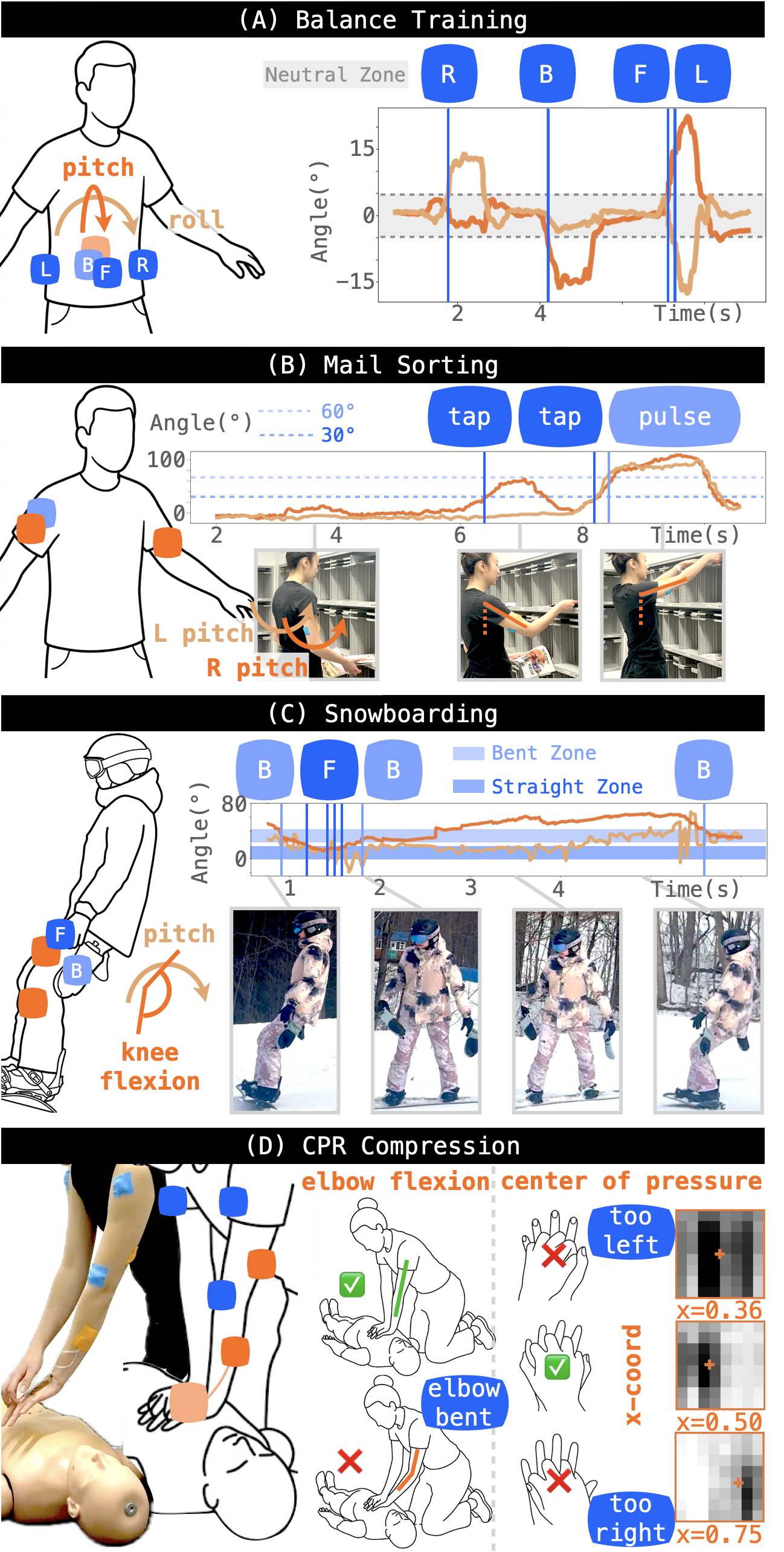}
  \vspace{-15pt}
  \caption{Task configurations based on prior works.}
  \label{fig:replicated_applications}
\end{figure}

We first evaluate whether the platform’s configuration model supports a meaningful range of training scenarios.
In Fig.~\ref{fig:replicated_applications} and Table~\ref{tab:example-task-configs}, we illustrate with 4 examples, 1 from each domain in Table~\ref{tab:related-work-domains}.

In balance training~\cite{vibrotactile-for-standing}, Fig.~\ref{fig:replicated_applications}(A), spatially mapped cues around the waist support immediate correction of trunk sway, following a common wearable rehabilitation design (e.g., belt-like systems~\cite{vibrotactile-for-standing}). 
The task used a neutral-standing calibration for reference.

In ergonomics~\cite{inertial-work-activities-review}, Fig.~\ref{fig:replicated_applications}(B), we adapted prior reminder-style feedback for reducing adverse upper-arm posture while sorting mail, where concurrent cues can help workers maintain attention on posture despite high cognitive demand~\cite{imu-vibrotacrtile-ergonomics-intervention}. 
The task used an arms-hanging calibration and two elevation limits (30\textdegree{} and 60\textdegree{}).

In snowboarding, Fig.~\ref{fig:replicated_applications}(C), where real-time coaching during movement is especially difficult~\cite{Wearable-automatic-feedback-snowboard}, two coauthors had a goal of becoming more aware of knee bend, which helps with balance and movements.
Two authors used the system themselves, using 2 separate phones, for the entire 90-minute snowboarding session, without interference.
With boots strapped in, calibration included knee flexion movements for axis identification followed by a natural standing reference.
Standing-gated (using thigh orientation) cues avoided triggers on lifts and after falls.

In cardiopulmonary resuscitation (CPR) training, Fig.~\ref{fig:replicated_applications}(D), prior systems~\cite{CPR-band, CPR-glove} suggest both promise and limitation: vibrotactile feedback can reduce visual distraction, but more complex cue vocabularies can increase learning burden, motivating more personalizable mappings. 
We therefore complement mannequin-based feedback with cues for elbow straightness and top-hand centering; calibration included elbow flexion and pronation movements for axis identification, a straight-elbow reference with shoulders over the hands, and a tactile baseline.
This training exercise later served as the preconfigured task in our end-user configuration study.

\section{Practitioner Elicitation \& Co-Config Study}\label{sec:practioner-study}
Sec.~\ref{sec:applications} demonstrated the breadth of applications that \thePlatform{} can instantiate.
We then evaluate whether movement-domain practitioners, experts in training and feedback, can map grounded scenarios from practice onto that same configuration model.
We conducted an IRB-approved practitioner elicitation and co-configuration study with practitioners to elicit application ideas and training contexts from their own practice, co-configure selected tasks on the platform, and examine what these uses revealed about workflow integration, fit, and reconfiguration needs.

\begin{table}[b]
\centering
\vspace{-10pt}
\caption{Practitioner Expert Participants.}
\label{tab:practitioner-background}
\small
\setlength{\tabcolsep}{3pt}
\begin{tabularx}{\columnwidth}{@{}p{0.05\columnwidth} p{0.3\columnwidth} X@{}}
\toprule
\textbf{ID} & \textbf{Role \& Years Exp.} & \textbf{Main Population Served / Level} \\
\midrule
E1 & athletic trainer, 22 & NCAA DI athletes, primarily in track \& field \\
\midrule
E2 & gymnastics coach, 30  & beginner \& recreational elementary-aged children and college students \\
\midrule
E3 & ballet instructor, 5  & beginner \& intermediate children (ages 6--12) and college club dancers \\
\midrule
E4 & athletic trainer, 2  & NCAA DI athletes, primarily in soccer \\
\midrule
E5 & wrestling coach, 15  & NCAA DI and Olympic hopeful wrestlers \\
\midrule
\multirow[t]{2}{*}{E6}
  & \makecell[tl]{pediatric physical\\therapist, 22} & elementary school students with individualized education programs (IEPs) \\
  & ballet instructor, 10 & recreational adults \\
\bottomrule

\end{tabularx}
\end{table}
\subsection{Participants \& Study Protocol}
We recruited 6 sensorimotor training practitioners via email and word of mouth. 
Table~\ref{tab:practitioner-background} summarizes their roles, experience (average years=16, std=10), and primary populations served. 
The participants worked across diverse populations and goals, ranging from recreational beginners and elementary-aged children to collegiate and Olympic-hopeful athletes, and from post-surgical rehabilitation and injury prevention to foundational technique and skill development.
Each study lasted about 90 minutes and took place either in the participant's training facility or in an experiment room, at their convenience. 
Participants were compensated US\$40.

The study began with consent and a semi-structured interview about recent movement-training tasks, important observations, typical feedback practices, and current challenges.
Participants then examined the physical stickies, received a walkthrough of the system capabilities, discussed initial impressions, concerns, sensing metrics of interest, and relevant vibrotactile feedback, and briefly tried feedback stickies on themselves.
Next, each practitioner worked with the researcher to co-configure one task drawn from their own practice.
Task and feedback-policy creation were researcher-supported: participants described the training goal they had in mind, and the researcher collaborated with them to shape the sensing and feedback logic.
In contrast, participants themselves carried out the setup workflow, including cutting tapes, placing stickies, choosing feedback locations, and completing calibration.
Participants then experienced the configured feedback system in practice.
We then revisited how this use aligned or misaligned with their earlier understanding of the platform and the selected training goals, where they saw opportunities for workflow integration, and what they would want improved or changed.
Finally, participants completed a brief questionnaire.
Studies were audio-recorded and transcribed.
Two authors independently inductively coded the transcripts, then met to compare interpretations, discuss discrepancies, and consolidate recurring themes across participants.

\subsection{Co-Configured Tasks}
\begin{figure}[b]
\vspace{-10pt}
  \includegraphics[width=\columnwidth]{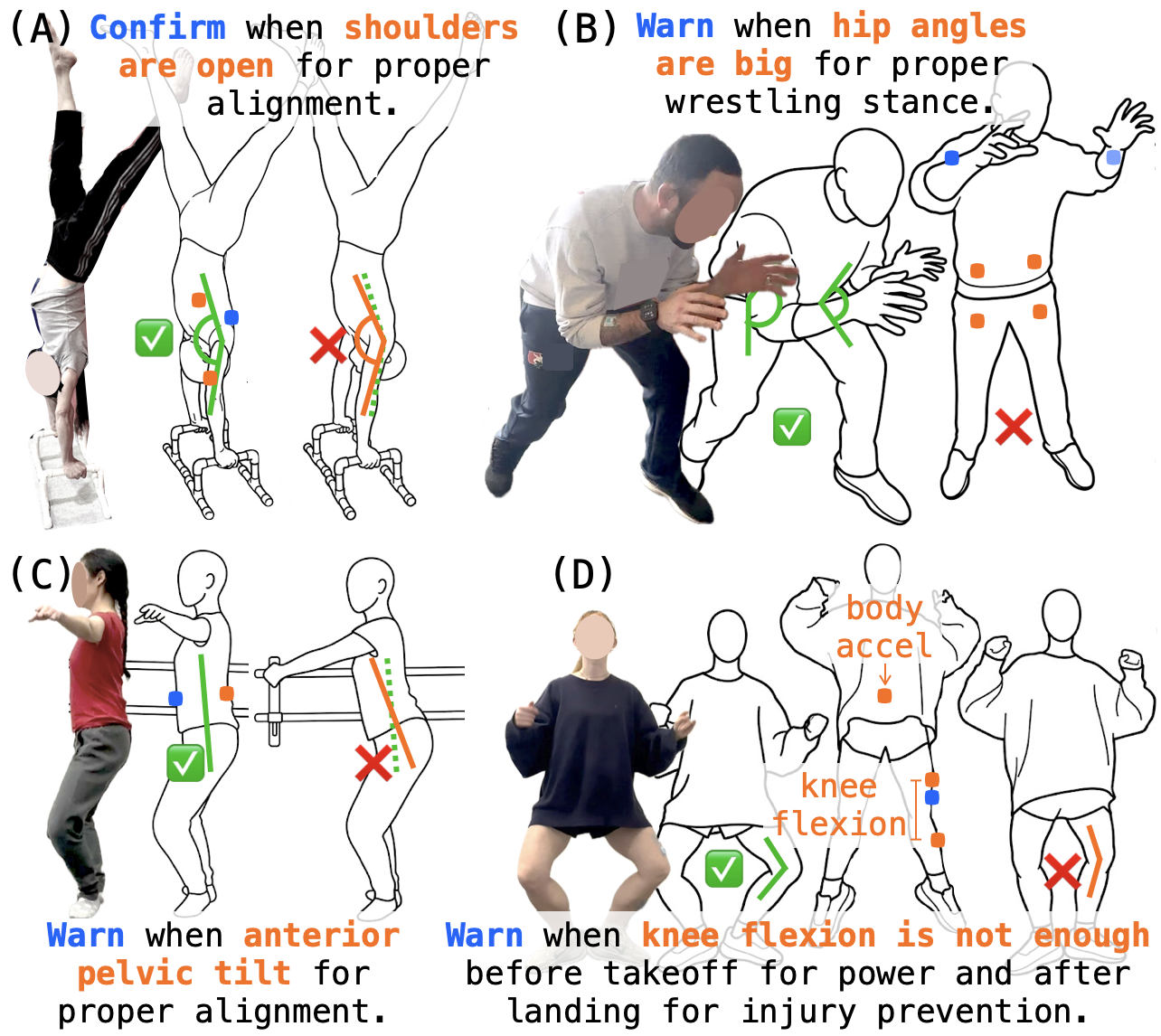}
  \vspace{-18pt}
  \caption{Configurations by (A) E2, the gymnastics coach, (B) E5, the wrestling coach, (C) E6, the ballet instructor, and (D) E3, the other ballet instructor.}
  \label{fig:expert-applications}
\end{figure}

All practitioners quickly identified multiple grounded scenarios from their own practice, but due to time constraints, each then selected one to co-configure on the platform, as shown in Fig.~\ref{fig:expert-applications}.

The 2 athletic trainers (ATs) configured different tasks within the \textbf{ACL post-surgical rehabilitation} broader context, detailed in Fig.~\ref{fig:walkthrough}.
E1 focused on a \textbf{flexion-limit warning}, where knee flexion had to remain within a range that changes across recovery stages.
E4 focused on a \textbf{full-extension reward}, where restoring terminal knee extension can be difficult to monitor continuously.

The gymnastics coach (E2), ballet instructors (E3, E6), and wrestling coach (E5) chose technique-oriented scenarios centered on alignment and movement quality, foundational skills practiced repeatedly from beginners to professionals.
E2 focused on \textbf{handstand shoulder alignment}: maintaining an open shoulder position matters, and subtle deviations are not always easy for learners to feel or see when upside down.
For ballet, E3 focused on \textbf{jump mechanics}, using feedback to cue a deeper knee bend before takeoff for power and after landing for safety, while E6 focused on \textbf{pelvic orientation for alignment}. 
E5 chose \textbf{stance-in-motion training}, where maintaining stance is fundamental even for advanced athletes, especially under fatigue.
Feedback alerted the athlete when they came out of the desired stance during repeated drills.

Although these applications relied on simple metrics (1--2 sensed variables) and simple rule logic, they covered a broad range of meaningful training situations.
But the current platform scope is not exhaustive: E3 initially wanted to author a task around knee valgus, which the metric library does not support, though feasible using IMUs.
Nevertheless, across configured cases, the pattern remained consistent: practitioners identified a meaningful movement variable, defined good or bad conditions, and used immediate on-body feedback to support awareness, reminders, and correction, suggesting that the platform's configuration model already captures a meaningful subset of real training needs.

\subsection{Qualitative Findings}
\subsubsection{\textbf{Practitioners reasoned about training goals in terms of simple sensed variables and cueing needs.}}
Practitioners described meaningful training goals, both in configured applications and additional envisioned uses, not as abstract "better movement", but in terms of a few concrete sensed variables and cueing needs. 
They rated expressing a desired training task on the platform as moderately easy (median = 4/5, IQR = 0.75).

\underline{Simple sensed variables.}
Rather than requiring an exhaustive representation of the body, practitioners focused on a small set of interpretable quantities tied to the body region that mattered for the drill or exercise, such as joint angles, limb orientation, or, in some envisioned cases, pressure and contact symmetry. 
This lets them specify what body-state mattered, what counted as good or bad, when feedback should occur, and where it should be felt.
These descriptions often already aligned with the platform’s task-, policy-, and session-level abstractions.
For example, the wrestling coach described stance training in terms of athlete-specific “parameters” before seeing how the platform supports parameter tuning.

\underline{Cueing needs.}
Practitioners also described feedback in similarly concrete terms. 
All 6 framed it primarily as a reminder, warning, confirmation, or, in one athletic training case, a reward-like signal when a desired position was reached. 
For example, one AT described vibrotactile feedback as a way to cue where someone should be contracting while “not actually doing the contraction for them” (E1), while the other AT noted that some athletes may respond well to feedback that marks reaching a target position (E4). 
These responses suggest that feedback authoring needs to support different communicative intents, not only different thresholds and patterns. 
5 practitioners also explicitly preferred placing feedback near the relevant body region, framing it as localized biofeedback (N=2). 
All 6 additionally expressed interest in seeing the underlying metrics visually, suggesting that sparse on-body cues and other feedback modalities may play complementary roles in practice.

\subsubsection{\textbf{Practitioners identified workflow opportunities alongside clear boundaries and tensions.}}
After trying the configured feedback in use, practitioners reflected on where the platform might fit into their existing workflows and where it might not.

\underline{Workflow opportunities.}
One recurring opportunity was limited trainer time and attention, both in group teaching contexts such as ballet and gymnastics and in sports medicine, where even well-supported collegiate athletes do not receive unlimited one-on-one attention.
Practitioners (N=4) saw value in feedback that could extend coaching beyond moments of direct supervision.
Another opportunity was helping learners notice movement qualities they often miss on their own.
5 practitioners explicitly discussed limited body awareness caused by either fatigue (for advanced performers) or difficulty in feeling the body in the moment (for beginners).
Vibrotactile feedback was seen as well-suited to these situations because it could provide immediate cues during movement rather than delayed correction afterward.
As E6 contrasted its immediacy with verbal correction like ``oh, no, don’t do it that way”.

\underline{Boundaries of fit.}
Practitioners identified clear ill-fit conditions.
They saw weaker fits for less structured or interpersonal contexts, such as pediatric therapy sessions or live wrestling with a partner.
2 practitioners also pointed to cases where feedback based on a small set of metrics and rules was a weaker fit, such as expressive solo dance or settings where the judgment was more context-dependent.
E6 similarly contrasted her two roles by describing the platform as more immediately promising for adult ballet alignment work than for pediatric physical therapy, where goals can be broader, attention spans shorter, and movement patterns less stable.

\underline{Adoption tensions.}
The in-session trials also revealed tensions beyond task fit.
3 practitioners worried that too much numerical information could be distracting or overwhelming for certain learners.
Practitioners had mixed opinions on tape-based attachment: some suggested straps/bands (N=3) or integrated garments, despite tradeoffs in slipping and reconfiguration.
Others emphasized that the main barrier was less setup or calibration, which E4 found “pretty time-efficient,” and more the preparation work of authoring tasks and policies from scratch. As E4 put it, “When things are sort of pre-programmed, it becomes a little bit easier to use.” 
This suggests that starter templates or reusable configurations may be important for real-world adoption.
Practitioners also asked about sweat resistance, which is critical for athletic contexts.

\subsubsection{\textbf{Even within a single training application, deployment required personalized reconfiguration.}}
All practitioners described personalization and progression as routine parts of their training practice, and found our platform easy to reconfigure feedback placements (median = 4/5, IQR = 0), timing (median = 4/5, IQR = 1.5), and patterns (median = 4/5, IQR = 0.8).

\underline{Across people.} 
Practitioners discussed individual differences, including body size and proportions (N=5), body sensitivity (N=3), proprioceptive awareness (N=4), and skill level (N=5), and how learners interpreted and responded to feedback.
The gymnastics coach pointed out that ``everyone’s got a different body they’re working with''.
The wrestling coach similarly discussed that different weight classes call for different stance angles.

\underline{Changing needs within a person.} 
5 practitioners described changing needs over time.
In rehab, what counts as acceptable motion can change across recovery phases: athletes start with a narrow set of movements, then later progress to a broader range of activities.
Similarly, the overall goal may stay conceptually similar while the expected quality or cueing strategy change over time.
The wrestling coach pointed out that “the ability to have different parameters, even for the same athlete, with a different drill is really important.”

\section{End-User Configuration Study}
The practitioner study showed that the platform maps onto meaningful training goals. 
We now evaluate whether end users can complete the practical steps of first-time use, including setup, calibration, and within-task reconfiguration.
We conducted an end-user study, approved by the Institutional Review Board (IRB), using the preconfigured cardiopulmonary resuscitation (CPR) task from Sec.~\ref{sec:applications}.
Note, the study focused on first-time use of a preconfigured application and reconfiguration, not feedback efficacy.

\subsection{Participants \& Study Protocol}
We recruited 10 participants (5 self-identified as female, 5 as male, mean age=28, std age=3.4):
4 participants regularly wore wearables, 5 had prior experience with wearable vibration feedback.
CPR experience spanned no prior training (n=2), a little prior training (n=5), and formal training (n=3).  
Each study lasted about 1 hour in an experiment room on a university campus, with US\$20 compensation.

Each study began with a brief platform walkthrough covering the stickies, adhesives, tactile patch interpretation, power-on procedure, and wearable vibrotactile sensation. Participants then watched a short CPR instructional video\footnote{\vspace{-20pt}https://www.youtube.com/shorts/IQAnNRQ8W\_g} and opened the preconfigured CPR task, as shown in Fig.~\ref{fig:tactile-screenshots}.  
The task, detailed in Sec.~\ref{sec:applications}, provides feedback for (1) bent elbow, (2) top hand too left, and (3) top hand too right. The left-elbow flexion used the 2-DoF joint-angle pipeline, which made setup and calibration a relatively demanding first-time-use case, requiring both flexions and pronations.
Participants followed the app guidance to complete setup as independently as possible by placing the required sensing stickies, choosing how many feedback stickies to use and where to place them, and completing calibration. The researcher answered clarification questions throughout, but did not operate the app or attach stickies to participants' bodies.
After calibration, the researcher measured the left elbow angle at 15\textdegree{}, 45\textdegree{}, and 90\textdegree{} positions with a goniometer for comparison with the app estimates. 
Participants then performed a short CPR practice on a mannequin and adjusted parameters and vibration settings as needed. 
After participants had familiarized themselves with the current feedback setup, they completed a per-configuration questionnaire with three 7-point Likert items. 
Participants then completed one or more rounds of feedback reconfiguration, each followed by another short practice and the same questionnaire. 
During these rounds, they could change the number and placement of feedback stickies, vibration duration, and threshold values. 
At the end of the session, we repeated the goniometer check (except for one study in which the app was accidentally closed before the check and thus lost the calibration status) and administered an end-of-study questionnaire comprising 5-point ratings of the physical stickies, 7-point ratings about setup and personalization, the 10 standard SUS items on a 5-point scale, and open-ended questions.
\subsection{Results}\label{sec:end-user-study-results}

\subsubsection{\textbf{First-time users could complete setup and calibration, though calibration guidance needs improvement.}}
All participants completed setup \& calibration and proceeded to the task trials without the experimenter's hands-on assistance, although all asked questions, especially during sensing stickies attachment (e.g., where to place the sensing stickies) and the calibration (e.g., how to do the calibration movements) process. 
On the questionnaire, participants reported that they could set up the preconfigured CPR task mostly on their own (median = 5/7, IQR = 1) and felt confident using it again in the future (median = 6/7, IQR = 1).
Based on our app log, this first setup (from entering the setup screen to leaving for calibration) took 12.4 minutes (std=2.6) on average, including asking clarification questions and thinking about feedback placements.

The left-elbow flexion checks provide a feasibility check on user-led sensing placement and calibration. 
As shown in Fig.~\ref{fig:sensing-accuracy}, across the 15\textdegree{}, 45\textdegree{}, and 90\textdegree{} goniometer references, the app's readings suggest that participants could place the sensing stickies and calibrate them well enough to support the task.
The measurements showed participant-specific offsets and spread, especially at larger angles: signed bias decreased from +6.1\textdegree{} to +2.5\textdegree{}, while overall MAE remained similar (9.0\textdegree{} vs. 9.1\textdegree{}). 
These MAE values, from discrete handheld goniometer checks, should not be interpreted as the average error of the algorithm or the system (see the original 1-DoF and 2-DoF joint-angle estimation papers for average errors~\cite{1dof-algo, 2dof-algo}).
Because the straight-arm reference depended on participant-performed calibration movements rather than a rigid external 0\textdegree{} posture, we interpret these measurements as evidence of feasible self-placement and calibration rather than precise absolute joint-angle estimation.
In the studies, practitioners and users inspected live metrics and adjusted thresholds during configuration, so task use does not rely on a single fixed global angle threshold.
Nevertheless, calibration execution was also the hardest part (N=5). Future work should improve calibration guidance through clearer pose references, video demonstrations, or improved calibration algorithms.

\begin{figure}[t]
  \includegraphics[width=\columnwidth]{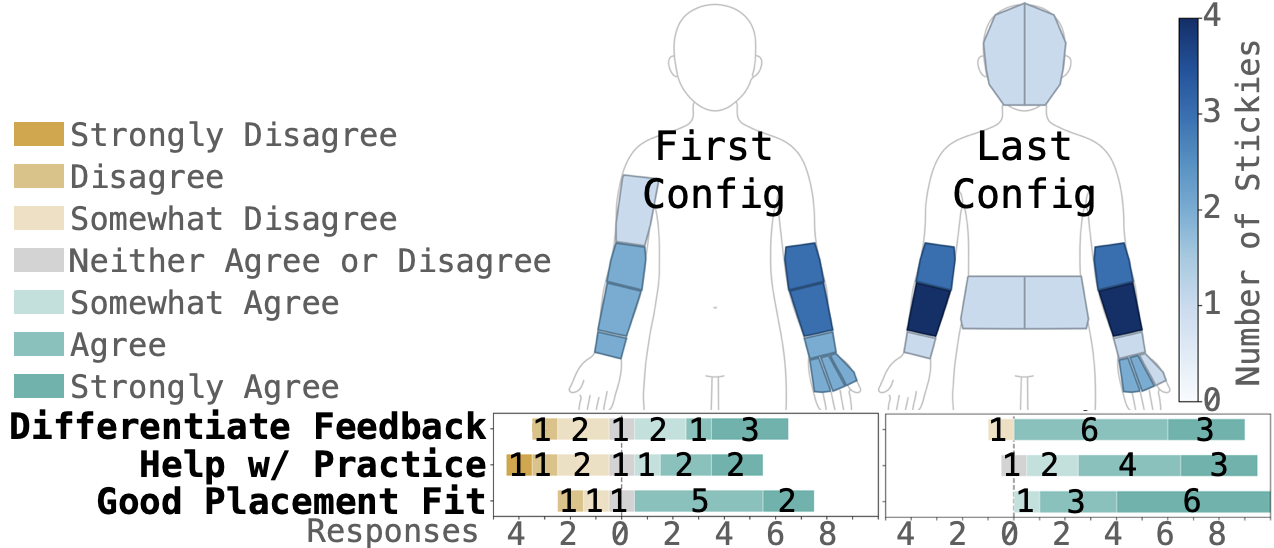}
  \caption{First and last end-user feedback configurations (as shown in the body maps) in the CPR task and their per-config ratings (as shown in the stacked bar charts).}
  \vspace{-16pt}
  \label{fig:end_user_results}
\end{figure}
\subsubsection{\textbf{Within-task personalization was easy and made a noticeable difference in use.}}
Participants found the task easy to personalize (median = 6/7, IQR = 1).
Two participants reported that their initial configuration already matched their desired setup and therefore did not make further changes; six produced two configurations, and two produced three. 
Due to the time constraints, 4 participants indicated they would have liked to explore additional configurations beyond what the session allowed.
Between the first and last configurations (see Fig.~\ref{fig:study-configurations} for all configurations), 4 of 10 participants changed the number of feedback stickies, and 5 of 10 changed at least one placement.
Configurations also shifted toward richer feedback layouts, from 3 one-sticky, 4 two-sticky, and 3 three-sticky starts to 1, 3, and 6 in final configurations, respectively.

Per-trial ratings, as shown in Fig.~\ref{fig:end_user_results}, improved from first to last configuration: median ratings 5 $\rightarrow$ 6 for telling apart feedback events, 5 $\rightarrow$ 6 for helping CPR practice, and 6 $\rightarrow$ 7 for placement fit. 
The first-config responses included neutral and disagree ratings that largely disappeared by the last config, suggesting that the value of reconfiguration was not only where participants ended up, but that the iteration made the feedback feel clearer and better fitted.

Final configurations remained diverse rather than converging to a single shared solution. 
Some participants preferred arm- and hand-based layouts, while others added feedback stickies to separate different events more clearly. 
Open-ended responses to what was useful to personalize help explain why this location flexibility mattered (N=10): participants appreciated putting the stickies "anywhere" (P2) and "wherever" (P9) that is "straightforward" (P1) and/or "at the corresponding area where the errors/mistakes are made"(P7) with explicit mentions of personalization reducing cognitive load (N=2). 
Interestingly, some participants wanted left \& right feedback locations (e.g., cheeks and waists) that were easy to distinguish, but others intentionally did not want to separate the two and instead preferred a generic indication for wrong hand position. 
These responses suggest that the main value was not convergence on a single best layout, but that iterative personalization, which may allow the system to work better for different users.

\subsubsection{\textbf{Users reported above-average usability and a positive overall experience.}}
The average SUS score was 78.75, between the 80th and 84th percentile and higher than average SUS scores~\cite{SUS}.
Responses suggest an overall positive experience with the system. Participants generally found the devices appealing and were willing to wear them (median = 4/5, IQR = 1), and rated the attachment (median = 4.5/5, IQR = 1.75) and detachment (median = 4/5, IQR = 0.75) process positively. Comfort while wearing was more mixed (median = 3.5/5, IQR = 1), mostly about the tactile patch.

\vspace{-5pt}
\section{Technical Characterization}\label{sec:technical-characterization}
Finally, we evaluate whether the underlying infrastructure is technically sufficient for untethered closed-loop deployment.
See Appendix~\ref{sec:appendix-characterization} for characterization details.
\\

\noindent\textbf{7.1 BLE Communication Latency.}
Closed-loop feedback requires low communication latency during movement, and our BLE characterization showed reasonable delay: \textbf{mean communication latency was 38.2 ms} (95\% CI: [37.8, 38.6]) \textbf{or 42.5 ms} (95\% CI: [41.9, 43.1]) \textbf{with the phone on the body or 5 m away}.
Adding the measured BLE signal-receive-to-feedback-dispatch time (mean = 5.7 ms, SD = 2.8, 95\% CI = [5.1, 6.3]), the total end-to-end latency is less than the typical human reaction time to vibrotactile stimuli~\cite{harrar2005simultaneity, peon2013reaction}.

\begin{figure}[t]    
\includegraphics[width=\columnwidth]{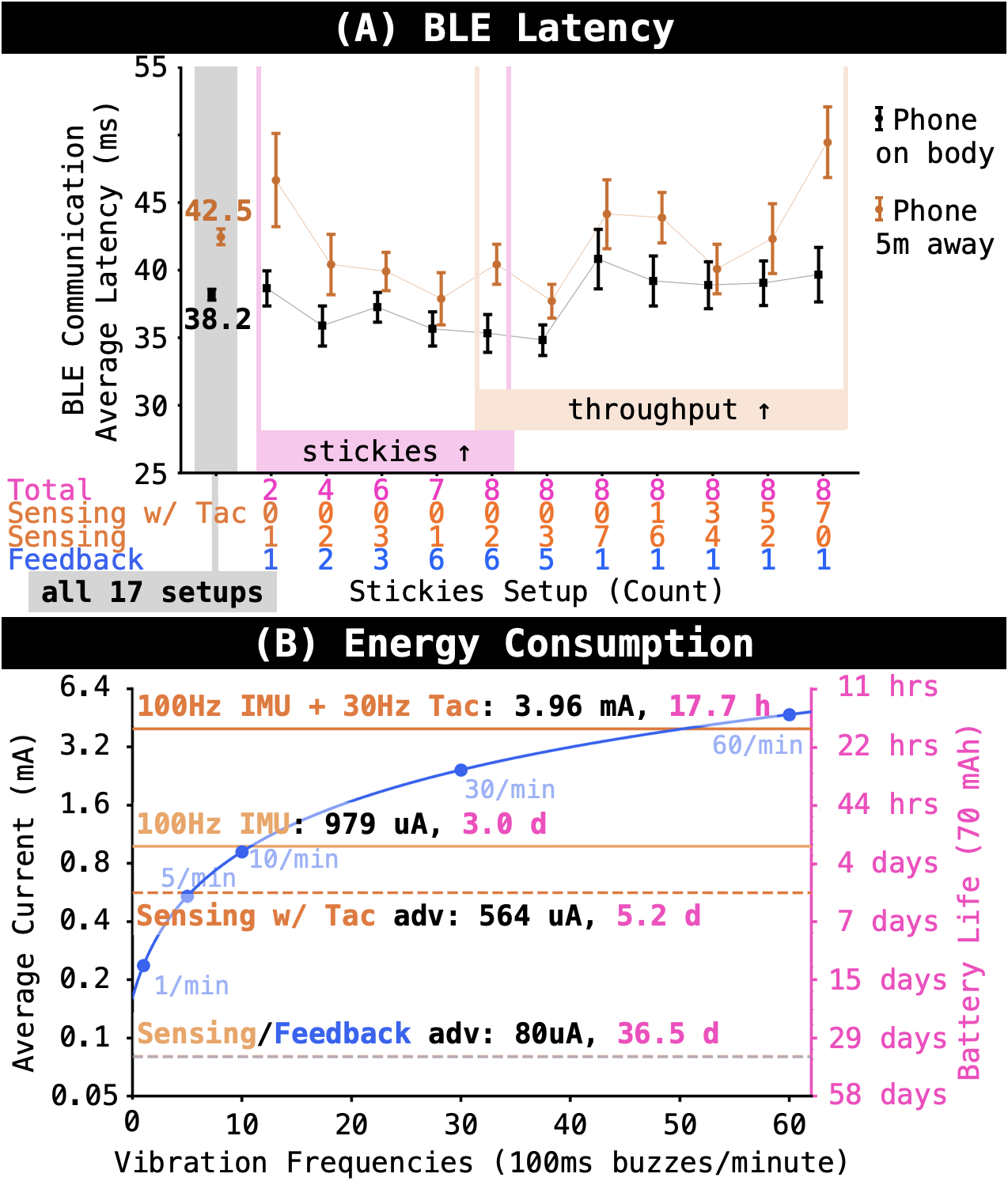}
\vspace{-15pt}
  \caption{(A) BLE Communication Latency: the leftmost column summarizes all setups that vary by stickies numbers and types. Error bars show 95\% confidence intervals. (B) Energy Consumption. ``Adv'' stands for BLE advertising.}
  \label{fig:characterization}
  \vspace{-10pt}
\end{figure}

As shown in Fig.~\ref{fig:characterization}(A), latency did not increase with the number of connected stickies: adding IMU-only sensing and feedback stickies left latency unchanged, while tactile-enabled sensing stickies caused only a modest increase.
\\

\noindent\textbf{7.2 BLE Packet Loss.}
Dropped sensing packets can disrupt real-time feedback timing and signal continuity, but our \textbf{measured 0.0015\% packet loss (32 packets) was negligible}.
These losses occurred only in the highest-throughput 5m-away configurations: sessions with 6 or 7 tactile-enabled sensing stickies and 2 other stickies, indicating stable transmission even near the upper limit.
\\

\noindent\textbf{7.3 Energy Consumption.}
Because the stickies are intended for untethered, prolonged wear, we measured current draw and estimated battery life using the selected 70 mAh LiPo battery:
\textbf{the most power-hungry setup still supported continuous full-day use, while more practical sparse-feedback usage could run for days}, as shown in Fig.~\ref{fig:characterization}(B).
Consistent with this, a kit of 14 stickies (6 feedback and 8 sensing) was used almost daily without recharging for 21 days; by day 21, all devices were still functioning.

Our current draw, with most modes drawing below 1 mA, is much smaller than that of similar prior systems, such as 60 mA for ErgoTac~\cite{Ergotac}, 66.6 mA for Dots~\cite{dots}, and 116 mA for Ariadne~\cite{Ariadne}. 
While these measurements are not from exactly matched conditions, they provide a context suggesting that our stickies support practical and prolonged deployment, enabled by efficient firmware.

\section{Discussion \& Future Directions}
Sensorimotor Stickies contributes a reconfigurable platform for closed-loop on-body sensorimotor training, evaluated in terms of breadth across tasks, practitioner-grounded configurations and fits, first-time setup and within-task configurations, and technical characterization for untethered deployment. However, toward broader real-world use, important challenges and opportunities remain.

\textbf{Richer authoring support could make the platform more immediately useful to a broader audience.} The current body-centered configuration model and companion app support threshold-based authoring across a wide range of contexts, but independent authoring requires familiarity with the configuration model that our studies did not assess. As E4 noted, for busy practitioners adopting a new technology, one way to reduce this learning burden in structured domains such as ACL rehabilitation may be template libraries that practitioners can directly use and adapt rather than build from scratch. A next step is therefore to host workshops in which attendees will learn the platform and author their own tasks, and, after open-sourcing, community contributions that can grow the current configuration model into a broader ecosystem.

\textbf{Real-world self-application will benefit from better support for calibration, attachment, and tactile sensing extension customization.} Our vision is not only that closed-loop sensorimotor feedback can be reconfigured, but that it can be patched onto the body and used in practice with little assistance. This requires better support for calibration guidance and sensing algorithms that tolerate non-expert placement and slipping (to be integrated into straps, as some practitioners suggested). This paper integrates existing sensing approaches with known limitations, and we hope the platform encourages researchers to improve this layer to make on-body sensing more accessible.  Mechanical robustness also needs consideration: while the PCBA stickies were robust, 1 connector on the tactile patch malfunctioned during repeated CPR compressions in the study, suggesting moving connectors further from active sensing areas. Tactile patches can also benefit from body- and task-specific size/shape customization.

\textbf{Sensorimotor Stickies points toward a broader platform for reconfigurable closed-loop interaction.} The present system intentionally focuses on a constrained but practical sensing-and-feedback stack that supports low power operation and sparse event-driven use. Meanwhile, the platform's architecture was designed with extensibility beyond the current scope.  The expansion port in the sensing sticky creates opportunities for additional sensing modalities. For the feedback sticky, the platform already supports preset patterns and app-defined temporal variations without firmware rewrite, while the infrastructure could support richer spatial patterns across multiple stickies. These directions could grow Sensorimotor Stickies from a focused platform into a broader family of configurable on-body sensing and feedback systems. Beyond on-skin use, stickies could be repurposed to be integrated into garments, braces, and/or objects.

\textbf{The open platform could grow through communities and AI agents.} Beyond the companion app, the same hardware and BLE infrastructure could support multiple authoring paths: task-specific UI for practitioners, lower-level programmability for developers, and shared community-authored configurations that make reconfiguration easier and broaden the platform's capabilities. With the rise of GenAI-assisted programming (e.g., “vibe coding”), we see potential for developing personalized apps and interaction logic on top of the exposed services. Looking further, Sensorimotor Stickies could serve as a physical layer through which AI agents access a user’s physical context for sensorimotor training and beyond. By making the reusable, we hope to also lower the barrier for future efficacy studies and longitudinal in-the-wild studies of how vibrotactile feedback shapes training and learning over time. Our snowboarding testing was an example in that direction, probing how the platform behaves when feedback must remain perceptible and reliable under clothing, motion, and sustained activity.

\section{Conclusion}
We presented \thePlatform{}, a reconfigurable platform for closed-loop on-body sensorimotor training that treats inertial and optional tactile sensing together with vibrotactile feedback not as task-specific one-off systems, but as a configurable stack where the underlying infrastructure stays constant while the configuration changes. 
By combining miniaturized adhesive sensing and feedback stickies, reusable low-power firmware and BLE infrastructure, and a body-centered configuration model realized through a companion mobile app, the platform supports reconfiguration across tasks, feedback policies, and feedback layouts without requiring the full stack to be rebuilt for each new task. 
Across configured tasks, a practitioner co-configuration study, an end-user configuration study, and technical characterization, our results suggest that this class of closed-loop wearable systems can be made more reusable, adaptable, and accessible. 
We hope \thePlatform{} lowers the barrier to building practical on-body closed-loop sensorimotor training aids and helps bring this class of systems closer to real-world use and study.

\begin{acks}
We sincerely thank the study participants and the reviewers. We thank Shu-Jung Han and Professor François Guimbretière for feedback on the study design. We also would like to thank the feedback and support from members of the SciFi Lab at Cornell and the Wearable Intelligence Lab at UW.
\end{acks}

\bibliographystyle{ACM-Reference-Format}
\bibliography{main}

@article{dots,
  title={Configurable, wearable sensing and vibrotactile feedback system for real-time postural balance and gait training: proof-of-concept},
  author={Xu, Junkai and Bao, Tian and Lee, Ung Hee and Kinnaird, Catherine and Carender, Wendy and Huang, Yangjian and Sienko, Kathleen H and Shull, Peter B},
  journal={Journal of neuroengineering and rehabilitation},
  volume={14},
  number={1},
  pages={102},
  year={2017},
  publisher={Springer}
}

@article{Ariadne,
  title={Open-source hardware and software platform for vibrotactile motion guidance},
  author={Rokhmanova, Nataliya and Martus, Julian and Faulkner, Robert and Fiene, Jonathan and Kuchenbecker, Katherine J},
  journal={Device},
  year={2025},
  publisher={Elsevier}
}

@article{TIKL,
  title={TIKL: Development of a wearable vibrotactile feedback suit for improved human motor learning},
  author={Lieberman, Jeff and Breazeal, Cynthia},
  journal={IEEE Transactions on Robotics},
  volume={23},
  number={5},
  pages={919--926},
  year={2007},
  publisher={IEEE}
}

@inproceedings{VibraForge,
  title={VibraForge: A Scalable Prototyping Toolkit For Creating Spatialized Vibrotactile Feedback Systems},
  author={Huang, Bingjian and Ren, Siyi and Luo, Yuewen and Cheng, Qilong and Cai, Hanfeng and Sang, Yeqi and Sousa, Mauricio and Dietz, Paul H and Wigdor, Daniel},
  booktitle={Proceedings of the 2025 CHI Conference on Human Factors in Computing Systems},
  pages={1--18},
  year={2025}
}

@inproceedings{Tape-Tics,
  title={Tape-Tics: A Flexible and Modular Vibrotactile Feedback System for Rapid Prototyping of Haptic Applications in Education},
  author={Paniagua, Carlos and Ota, Hiroki and Hirao, Yutaro and Perusquia-Hernandez, Monica and Uchiyama, Hideaki and Kiyokawa, Kiyoshi},
  booktitle={Proceedings of the Augmented Humans International Conference 2025},
  pages={92--104},
  year={2025}
}

@inproceedings{TactJam,
  title={TactJam: An end-to-end prototyping suite for collaborative design of on-body vibrotactile feedback},
  author={Wittchen, Dennis and Spiel, Katta and Fruchard, Bruno and Degraen, Donald and Schneider, Oliver and Freitag, Georg and Strohmeier, Paul},
  booktitle={Proceedings of the Sixteenth International Conference on Tangible, Embedded, and Embodied Interaction},
  pages={1--13},
  year={2022}
}

@article{vibrotactile-for-standing,
  title={Vibrotactile feedback for improving standing balance},
  author={Ballardini, Giulia and Florio, Valeria and Canessa, Andrea and Carlini, Giorgio and Morasso, Pietro and Casadio, Maura},
  journal={Frontiers in bioengineering and biotechnology},
  volume={8},
  pages={94},
  year={2020},
  publisher={Frontiers Media SA}
}

@inproceedings{tactile-for-snowboard,
  title={An investigation into the use of tactile instructions in snowboarding},
  author={Spelmezan, Daniel},
  booktitle={Proceedings of the 14th international conference on Human-computer interaction with mobile devices and services},
  pages={417--426},
  year={2012}
}

@inproceedings{vibrotactile-arm-motions-lessons,
  title={Lessons in using vibrotactile feedback to guide fast arm motions},
  author={Bark, Karlin and Khanna, Preeya and Irwin, Rikki and Kapur, Pulkit and Jax, Steven A and Buxbaum, Laurel J and Kuchenbecker, Katherine J},
  booktitle={2011 IEEE World Haptics Conference},
  pages={355--360},
  year={2011},
  organization={IEEE}
}

@article{Intercorporeal-biofeedback,
  title={Intercorporeal biofeedback for movement learning},
  author={Turmo Vidal, Laia and M{\'a}rquez Segura, Elena and Waern, Annika},
  journal={ACM Transactions on Computer-Human Interaction},
  volume={30},
  number={3},
  pages={1--40},
  year={2023},
  publisher={ACM New York, NY}
}

@article{review-real-time-biomechanical-feedback,
  title={Review of real-time biomechanical feedback systems in sport and rehabilitation},
  author={Hribernik, Matev{\v{z}} and Umek, Anton and Toma{\v{z}}i{\v{c}}, Sa{\v{s}}o and Kos, Anton},
  journal={Sensors},
  volume={22},
  number={8},
  pages={3006},
  year={2022},
  publisher={MDPI}
}

@article{vibrotacitle-ergonomic,
  title={A directional vibrotactile feedback interface for ergonomic postural adjustment},
  author={Kim, Wansoo and Garate, Virginia Ruiz and Gandarias, Juan M and Lorenzini, Marta and Ajoudani, Arash},
  journal={IEEE Transactions on Haptics},
  volume={15},
  number={1},
  pages={200--211},
  year={2021},
  publisher={IEEE}
}

@inproceedings{VibroBits,
  title={VibroBits: Visualising Body Centre of Pressure Using Vibrotactile Feedback for Sports Training},
  author={Weerasinghe, Chathura Nirmal and Abeywardhane, Ruchira and Kodikara, Uvindu and Abewickrema, Sasindu and Sabnis, Nihar and Mueller, Florian'Floyd' and Elvitigala, Don Samitha},
  booktitle={Proceedings of the First Annual Conference on Human-Computer Interaction and Sports},
  pages={1--4},
  year={2025}
}

@article{Musicjacket,
  title={Musicjacket—combining motion capture and vibrotactile feedback to teach violin bowing},
  author={Van Der Linden, Janet and Schoonderwaldt, Erwin and Bird, Jon and Johnson, Rose},
  journal={IEEE Transactions on Instrumentation and Measurement},
  volume={60},
  number={1},
  pages={104--113},
  year={2010},
  publisher={IEEE}
}

@inproceedings{sitting-posture-review,
  title={Sitting posture recognition and feedback: A literature review},
  author={Krauter, Christian and Angerbauer, Katrin and Sousa Calepso, Aim{\'e}e and Achberger, Alexander and Mayer, Sven and Sedlmair, Michael},
  booktitle={Proceedings of the 2024 CHI Conference on Human Factors in Computing Systems},
  pages={1--20},
  year={2024}
}

@misc{XSens,
  year = {2026},
  author = {Movella},
  title = {Xsens Products},
  howpublished = {\url{https://www.movella.com/products/xsens}}}

@misc{Vicon,
  year = {2026},
  author = {Vicon},
  title = {Vicon | Award Winning Motion Capture Systems},
  howpublished = {\url{https://www.vicon.com/}}}

@misc{OptiTrack,
  year = {2026},
  author = {OptiTrack},
  title = {OptiTrack - Motion Capture Systems},
  howpublished = {\url{https://optitrack.com/}}}

@article{OpenSim,
  title={OpenSim: open-source software to create and analyze dynamic simulations of movement},
  author={Delp, Scott L and Anderson, Frank C and Arnold, Allison S and Loan, Peter and Habib, Ayman and John, Chand T and Guendelman, Eran and Thelen, Darryl G},
  journal={IEEE transactions on biomedical engineering},
  volume={54},
  number={11},
  pages={1940--1950},
  year={2007},
  publisher={IEEE}
}

@inproceedings{toolkit-evaluation,
  title={Evaluation strategies for HCI toolkit research},
  author={Ledo, David and Houben, Steven and Vermeulen, Jo and Marquardt, Nicolai and Oehlberg, Lora and Greenberg, Saul},
  booktitle={Proceedings of the 2018 CHI conference on human factors in computing systems},
  pages={1--17},
  year={2018}
}

@inproceedings{VHP,
  title={VHP: vibrotactile haptics platform for on-body applications},
  author={Dementyev, Artem and Getreuer, Pascal and Kanevsky, Dimitri and Slaney, Malcolm and Lyon, Richard F},
  booktitle={The 34th Annual ACM Symposium on User Interface Software and Technology},
  pages={598--612},
  year={2021}
}

@inproceedings{Sound2Haptic,
  title={Sound2Haptic: A Toolkit for Portable Multi-Channel Haptic Integration Across Multiple Form Factors and Devices},
  author={Chin, Sam and Fitz-Gibbon, Emmie and Huang, Bingjian and Tims, Carter and Orzech, Gabrielle and Thoo, Yong-Joon and Paradiso, Joseph A},
  booktitle={Adjunct Proceedings of the 38th Annual ACM Symposium on User Interface Software and Technology},
  pages={1--4},
  year={2025}
}

@inproceedings{Hapticus,
  title={Hapticus: Exploring the Effects of Haptic Feedback and its Customization on Motor Skill Learning: Tactile, Haptic, and Somatosensory Approaches},
  author={Lee, Kyungyeon and Yang, Daniel S and Singh, Kriti and Nishida, Jun},
  booktitle={Proceedings of the 2025 CHI Conference on Human Factors in Computing Systems},
  pages={1--20},
  year={2025}
}

@article{Scene2Hap,
  title={Scene2Hap: Combining LLMs and physical modeling for automatically generating vibrotactile signals for full VR scenes},
  author={Jingu, Arata and AliAbbasi, Easa and Strohmeier, Paul and Steimle, J{\"u}rgen},
  journal={arXiv preprint arXiv:2504.19611},
  year={2025}
}

@article{Sensorimotor-Learn-VR,
  title={Learning and transfer of complex motor skills in virtual reality: a perspective review},
  author={Levac, Danielle E and Huber, Meghan E and Sternad, Dagmar},
  journal={Journal of neuroengineering and rehabilitation},
  volume={16},
  number={1},
  pages={121},
  year={2019},
  publisher={Springer}
}

@inproceedings{AdaptiveEMS,
  title={Adaptive Electrical Muscle Stimulation Improves Muscle Memory},
  author={Choudhary, Siya and Nith, Romain and Ho, Yun and Brooks, Jas and Guruvugari, Mithil and Lopes, Pedro},
  booktitle={Proceedings of the 2025 CHI Conference on Human Factors in Computing Systems},
  pages={1--11},
  year={2025}
}

@article{Skinlink,
  title={Skinlink: On-body construction and prototyping of reconfigurable epidermal interfaces},
  author={Ku, Pin-Sung and Huang, Kunpeng and Wang, Nancy and Ng, Boaz and Chu, Alicia and Kao, Hsin-Liu Cindy},
  journal={Proceedings of the ACM on Interactive, Mobile, Wearable and Ubiquitous Technologies},
  volume={7},
  number={2},
  pages={1--27},
  year={2023},
  publisher={ACM New York, NY, USA}
}

@article{Thermal-earring,
  title={Thermal earring: low-power wireless earring for longitudinal earlobe temperature sensing},
  author={Xue, Qiuyue Shirley and Liu, Yujia and Breda, Joseph and Springston, Mastafa and Iyer, Vikram and Patel, Shwetak},
  journal={Proceedings of the ACM on Interactive, Mobile, Wearable and Ubiquitous Technologies},
  volume={7},
  number={4},
  pages={1--28},
  year={2024},
  publisher={ACM New York, NY, USA}
}

@inproceedings{LilyPad,
  title={The LilyPad Arduino: using computational textiles to investigate engagement, aesthetics, and diversity in computer science education},
  author={Buechley, Leah and Eisenberg, Mike and Catchen, Jaime and Crockett, Ali},
  booktitle={Proceedings of the SIGCHI conference on Human factors in computing systems},
  pages={423--432},
  year={2008}
}

@inproceedings{Wiresens,
  title={Wiresens toolkit: An open-source platform towards accessible wireless tactile sensing},
  author={Murphy, Devin and Zhu, Junyi and Gadre, Akshay and Torralba, Antonio and Liang, Paul Pu and Matusik, Wojciech and Luo, Yiyue},
  booktitle={Proceedings of the Twentieth International Conference on Tangible, Embedded, and Embodied Interaction},
  pages={1--15},
  year={2026}
}

@article{CPR-glove,
  title={A Closed-Loop CPR Training Glove with Integrated Tactile Sensing and Haptic Feedback},
  author={Moon, Jaeyoung and Ma, Mingzhuo and Yang, Qifeng and Choi, Youjin and Hwang, Seokhyun and Burden, Samuel and Kim, Kyung-Joong and Luo, Yiyue},
  journal={arXiv preprint arXiv:2603.05793},
  year={2026}
}

@inproceedings{trigger-action-programming,
  title={Practical trigger-action programming in the smart home},
  author={Ur, Blase and McManus, Elyse and Pak Yong Ho, Melwyn and Littman, Michael L},
  booktitle={Proceedings of the SIGCHI conference on human factors in computing systems},
  pages={803--812},
  year={2014}
}

@article{customize-domain-specific-tools,
  title={Empowering end users to customize their smart environments: model, composition paradigms, and domain-specific tools},
  author={Desolda, Giuseppe and Ardito, Carmelo and Matera, Maristella},
  journal={ACM Transactions on Computer-Human Interaction (TOCHI)},
  volume={24},
  number={2},
  pages={1--52},
  year={2017},
  publisher={ACM New York, NY, USA}
}

@article{smartfit-rule-editor,
  title={A visual language and interactive system for end-user development of internet of things ecosystems},
  author={Barricelli, Barbara Rita and Valtolina, Stefano},
  journal={Journal of Visual Languages \& Computing},
  volume={40},
  pages={1--19},
  year={2017},
  publisher={Elsevier}
}

@article{SUS,
  title={The system usability scale: past, present, and future},
  author={Lewis, James R},
  journal={International Journal of Human--Computer Interaction},
  volume={34},
  number={7},
  pages={577--590},
  year={2018},
  publisher={Taylor \& Francis}
}

@inproceedings{flexible-wearables-physical-training,
  title={Exploring opportunities for flexible wearables to support physical training},
  author={Stawarz, Katarzyna and Everitt, Aluna and Bowen, Judy},
  booktitle={Proceedings of the 2025 ACM Designing Interactive Systems Conference},
  pages={2156--2170},
  year={2025}
}

@article{attentional-focus,
  title={The automaticity of complex motor skill learning as a function of attentional focus},
  author={Wulf, Gabriele and McNevin, Nancy and Shea, Charles H},
  journal={The Quarterly Journal of Experimental Psychology Section A},
  volume={54},
  number={4},
  pages={1143--1154},
  year={2001},
  publisher={SAGE Publications Sage UK: London, England}
}

@inproceedings{kinematic-vibrotactile-mapping,
  title={Motor learning using a kinematic-vibrotactile mapping targeting fundamental movements},
  author={McDaniel, Troy and Goldberg, Morris and Villanueva, Daniel and Viswanathan, Lakshmie Narayan and Panchanathan, Sethuraman},
  booktitle={Proceedings of the 19th ACM international conference on Multimedia},
  pages={543--552},
  year={2011}
}

@incollection{frequent-feedback-degrade-learning,
  title={Frequent augmented feedback can degrade learning: Evidence and interpretations},
  author={Schmidt, Richard A},
  booktitle={Tutorials in motor neuroscience},
  pages={59--75},
  year={1991},
  publisher={Springer}
}

@article{agumented-feedback-motor-learning,
  title={Augmented visual, auditory, haptic, and multimodal feedback in motor learning: a review},
  author={Sigrist, Roland and Rauter, Georg and Riener, Robert and Wolf, Peter},
  journal={Psychonomic bulletin \& review},
  volume={20},
  number={1},
  pages={21--53},
  year={2013},
  publisher={Springer}
}

@inproceedings{sports-wearables-design-space,
  title={The design space of wearables for sports and fitness practices},
  author={Turmo Vidal, Laia and Zhu, Hui and Waern, Annika and M{\'a}rquez Segura, Elena},
  booktitle={Proceedings of the 2021 CHI Conference on Human Factors in Computing Systems},
  pages={1--14},
  year={2021}
}

@article{Ergotac,
  title={Ergotac: A tactile feedback interface for improving human ergonomics in workplaces},
  author={Kim, Wansoo and Lorenzini, Marta and Kap{\i}c{\i}o{\u{g}}lu, Ka{\u{g}}an and Ajoudani, Arash},
  journal={IEEE Robotics and Automation Letters},
  volume={3},
  number={4},
  pages={4179--4186},
  year={2018},
  publisher={IEEE}
}

@article{inertial-work-activities-review,
  title={Evidence for the effectiveness of feedback from wearable inertial sensors during work-related activities: A scoping review},
  author={Lee, Roger and James, Carole and Edwards, Suzi and Skinner, Geoff and Young, Jodi L and Snodgrass, Suzanne J},
  journal={Sensors},
  volume={21},
  number={19},
  pages={6377},
  year={2021},
  publisher={MDPI}
}

@article{wearable-occupational-review,
  title={Advancing occupational medicine through wearable technology: A review of sensor systems for biomechanical risk assessment and work-related musculoskeletal disorder prevention},
  author={Babangida, Abubakar A and Caraballo-Arias, Yohama and Decataldo, Francesco and Violante, Francesco Saverio},
  journal={ACS sensors},
  volume={10},
  number={8},
  pages={5410--5432},
  year={2025},
  publisher={ACS Publications}
}

@article{vibrotactile-sports-review,
  title={Vibrotactile feedback as a tool to improve motor learning and sports performance: a systematic review},
  author={Van Breda, Eric and Verwulgen, Stijn and Saeys, Wim and Wuyts, Katja and Peeters, Thomas and Truijen, Steven},
  journal={BMJ open sport \& exercise medicine},
  volume={3},
  number={1},
  year={2017},
  publisher={BMJ Publishing Group Ltd}
}

@incollection{vibrotactile-deign-music,
  title={Design of vibrotactile feedback and stimulation for music performance},
  author={Giordano, Marcello and Sullivan, John and Wanderley, Marcelo M},
  booktitle={Musical haptics},
  pages={193--214},
  year={2018},
  publisher={Springer}
}

@article{vibrotactile-rehab-review,
  title={Vibrotactile-based rehabilitation on balance and gait in patients with neurological diseases: A systematic review and metanalysis},
  author={De Angelis, Sara and Princi, Alessandro Antonio and Dal Farra, Fulvio and Morone, Giovanni and Caltagirone, Carlo and Tramontano, Marco},
  journal={Brain sciences},
  volume={11},
  number={4},
  pages={518},
  year={2021},
  publisher={MDPI}
}

@inproceedings{GuiDance,
  title={GuiDance: Wearable technology applied to guided dance},
  author={Camarillo-Abad, H{\'e}ctor M and Sandoval, Mar{\'\i}a Gabriela and S{\'a}nchez, J Alfredo},
  booktitle={Proceedings of the 7th Mexican Conference on Human-Computer Interaction},
  pages={1--8},
  year={2018}
}

@inproceedings{seated-vibrotactile-garment,
  title={Sensory garments with vibrotactile feedback for monitoring and informing seated posture},
  author={Barone, Vincent J and Yuen, Michelle C and Kramer-Boniglio, Rebecca and Sienko, Kathleen H},
  booktitle={2019 2nd IEEE International Conference on Soft Robotics (RoboSoft)},
  pages={391--397},
  year={2019},
  organization={IEEE}
}

@inproceedings{GymSoles,
  title={GymSoles: Improving squats and dead-lifts by visualizing the user's center of pressure},
  author={Elvitigala, Don Samitha and Matthies, Denys JC and David, L{\"o}ic and Weerasinghe, Chamod and Nanayakkara, Suranga},
  booktitle={Proceedings of the 2019 CHI Conference on Human Factors in Computing Systems},
  pages={1--12},
  year={2019}
}

@article{CPR-band,
  title={The development and accuracy assessment of a wearable cardiopulmonary resuscitation real-time audio-visual feedback device},
  author={Ma, Wenwen and Liu, Enze and Xiao, Landan and Song, Yuanwen and Zhou, Liangyuan and Zhang, Chen and Deng, Huisheng},
  journal={Resuscitation},
  volume={211},
  pages={110602},
  year={2025},
  publisher={Elsevier}
}

@article{needle-wristband,
  title={Multiactuator haptic feedback on the wrist for needle steering guidance in brachytherapy},
  author={Rossa, Carlos and Fong, Jason and Usmani, Nawaid and Sloboda, Ronald and Tavakoli, Mahdi},
  journal={IEEE Robotics and Automation Letters},
  volume={1},
  number={2},
  pages={852--859},
  year={2016},
  publisher={IEEE}
}

@article{yoga-vibrotactile,
  title={Effects of vibrotactile feedback on yoga practice},
  author={Islam, Md Shafiqul and Lee, Sang Won and Harden, Samantha M and Lim, Sol},
  journal={Frontiers in Sports and Active Living},
  volume={4},
  pages={1005003},
  year={2022},
  publisher={Frontiers Media SA}
}

@article{imu-vibrotacrtile-ergonomics-intervention,
  title={A wearable sensor system for physical ergonomics interventions using haptic feedback},
  author={Lind, Carl Mikael and Diaz-Olivares, Jose Antonio and Lindecrantz, Kaj and Eklund, J{\"o}rgen},
  journal={Sensors},
  volume={20},
  number={21},
  pages={6010},
  year={2020},
  publisher={MDPI}
}

@inproceedings{Wearable-automatic-feedback-snowboard,
  title={Wearable automatic feedback devices for physical activities},
  author={Spelmezan, Daniel and Schanowski, Adalbert and Borchers, Jan},
  booktitle={Proceedings of the fourth international conference on body area networks},
  pages={1--8},
  year={2009}
}

@article{vqf,
  title={VQF: Highly accurate IMU orientation estimation with bias estimation and magnetic disturbance rejection},
  author={Laidig, Daniel and Seel, Thomas},
  journal={Information Fusion},
  volume={91},
  pages={187--204},
  year={2023},
  publisher={Elsevier}
}

@article{1dof-algo,
  title={Robust plug-and-play joint axis estimation using inertial sensors},
  author={Olsson, Fredrik and Kok, Manon and Seel, Thomas and Halvorsen, Kjartan},
  journal={Sensors},
  volume={20},
  number={12},
  pages={3534},
  year={2020},
  publisher={MDPI}
}

@article{2dof-algo,
  title={Self-calibrating magnetometer-free inertial motion tracking of 2-DoF joints},
  author={Laidig, Daniel and Weygers, Ive and Seel, Thomas},
  journal={Sensors},
  volume={22},
  number={24},
  pages={9850},
  year={2022},
  publisher={MDPI}
}

@inproceedings{peon2013reaction,
  title={Reaction times to constraint violation in haptics: comparing vibration, visual and audio stimuli},
  author={Peon, Adrian Ramos and Prattichizzo, Domenico},
  booktitle={2013 world haptics conference (WHC)},
  pages={657--661},
  year={2013},
  organization={IEEE}
}

@article{harrar2005simultaneity,
  title={Simultaneity constancy: detecting events with touch and vision},
  author={Harrar, Vanessa and Harris, Laurence R},
  journal={Experimental Brain Research},
  volume={166},
  number={3},
  pages={465--473},
  year={2005},
  publisher={Springer}
}

@inproceedings{aveni2025generative,
  title={Generative Trigger-Action Programming with Ply},
  author={Aveni, Timothy J and Mor, Hila and Fox, Armando and Hartmann, Bj{\"o}rn},
  booktitle={Proceedings of the 38th Annual ACM Symposium on User Interface Software and Technology},
  pages={1--17},
  year={2025}
}

\appendix

\begin{figure}[b]
\includegraphics[width=\columnwidth]{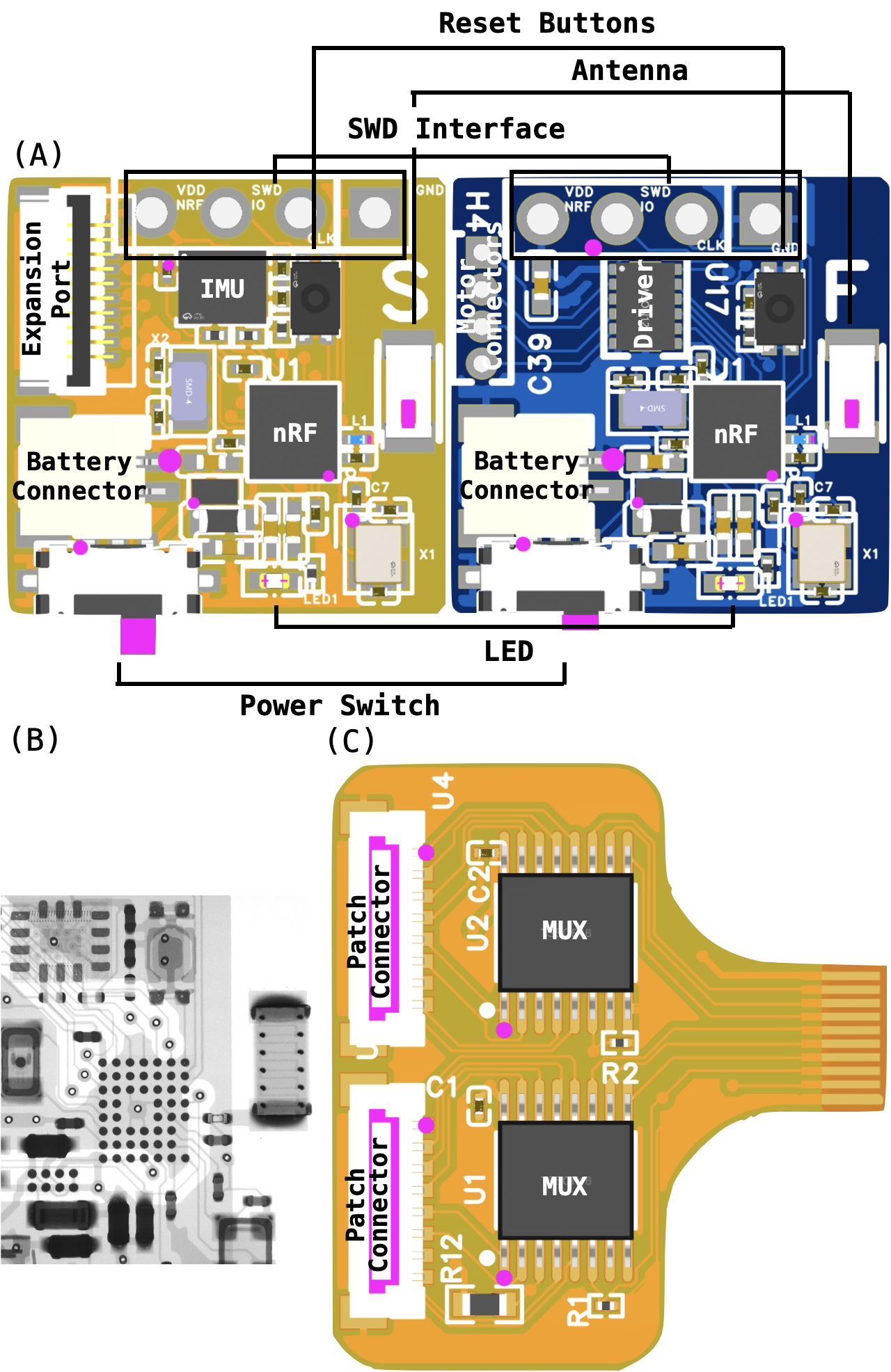}
  \caption{(A) Sensing and Feedback PCBs. (B) A partial X-ray scan of the sensing sticky. (C) Tactile Extension FPCB.}
  \label{fig:pcb-details}
\end{figure}

\section{Implementation Details}\label{sec:appendix-implementation}
This section provides additional implementation details for the wearable hardware, embedded firmware and BLE services, companion app and runtime, and sensing and calibration pipelines introduced in Sec.~\ref{sec:platform-overview}.
\renewcommand{\arraystretch}{1.08}
\setlength{\tabcolsep}{2.6pt}
\newcolumntype{L}{>{\raggedright\arraybackslash}X}
\begin{table}[b]
\centering
\caption{Component cost breakdown for the sensing board.}
\vspace{-10pt}
\label{tab:sensing_components_cost}
\small
{\hyphenpenalty=10000\exhyphenpenalty=10000
\begin{tabularx}{\columnwidth}{@{}L >{\raggedright\arraybackslash}p{0.12\columnwidth} >{\raggedright\arraybackslash}p{0.16\columnwidth}@{}}
\toprule
\textbf{Electrical Component} & \textbf{Retail (\$)} & \textbf{Wholesale (1k, \$)} \\
\midrule
MCU (NRF52832-CIAA-R) & 3.35 & 2.86 \\
IMU (BMI323) & 3.27 & 2.02 \\
Chip Antenna (ANT3216LL00R2400A) & 0.44 & 0.20 \\
Voltage Regulator (TPS62743YFPR) & 1.50 & 0.77 \\
Connectors (SM02B-SRSS-TB(LF)(SN), FH34SRJ-10S-0.5SH(50)) & 1.23 & 0.76 \\
Switches (GT-TC194A-H0055-L1, MK-11C04-G013) & 0.74 & 0.23 \\
Oscillators (X201632MKB4SI, FC-12M32.768KHZ9PF20PPM) & 1.16 & 0.51 \\
Others (resistors, capacitors, inductors, LED) & 3.16 & 0.40 \\
\addlinespace[2pt]
\textbf{Total} & \textbf{14.85} & \textbf{7.75} \\
\bottomrule
\end{tabularx}
}
\end{table}

\begin{table}[b]
\centering
\caption{Component cost breakdown for the feedback board.}
\vspace{-10pt}
\label{tab:feedback_components_cost}
\small
{\hyphenpenalty=10000\exhyphenpenalty=10000
\begin{tabularx}{\columnwidth}{@{}L >{\raggedright\arraybackslash}p{0.12\columnwidth} >{\raggedright\arraybackslash}p{0.16\columnwidth}@{}}
\toprule
\textbf{Electrical Component} & \textbf{Retail (\$)} & \textbf{Wholesale (1k, \$)} \\
\midrule
MCU (NRF52832-CIAA-R) & 3.35 & 2.86 \\
Motors (VC0720B015F) & 3.48 & 2.42 \\
Motor Driver (DRV8836DSSR) & 1.95 & 1.03 \\
Chip Antenna (ANT3216LL00R2400A) & 0.44 & 0.20 \\
Voltage Regulator (TPS62743YFPR) & 1.50 & 0.77 \\
Connectors (HB-PH3-25414PB2GOP, 1271WV-4P) & 0.03 & 0.01 \\
Switches (GT-TC194A-H0055-L1, MK-11C04-G013) & 0.74 & 0.23 \\
Oscillators (X201632MKB4SI, FC-12M32.768KHZ9PF20PPM) & 1.16 & 0.51 \\
Others (resistors, capacitors, inductors, LED) & 2.86 & 0.40 \\
\addlinespace[2pt]
\textbf{Total} & \textbf{15.51} & \textbf{8.43} \\
\bottomrule
\end{tabularx}
}
\end{table}
\subsection{Wearable Hardware}
Across the physical implementation, we prioritize miniaturization together with accessibility and manufacturability rather than pursuing the smallest possible construction.
\begin{table*}[t]
\centering
\caption{BLE GATT services and characteristics used by the stickies. \texttt{u8}, \texttt{u16}, \texttt{u32}, and \texttt{i16} denote unsigned/signed integer field widths. RTT stands for round-trip time.}
\vspace{-12pt}
\label{tab:ble_gatt}
\small
\setlength{\tabcolsep}{3pt}
\renewcommand{\arraystretch}{1.08}
\begin{tabularx}{\textwidth}{@{}
>{\raggedright\arraybackslash}p{0.14\textwidth}
>{\raggedright\arraybackslash}p{0.17\textwidth}
>{\raggedright\arraybackslash}p{0.09\textwidth}
>{\raggedright\arraybackslash}p{0.13\textwidth}
X@{}}
\toprule
\textbf{Service} & \textbf{Characteristic} & \textbf{UUID} & \textbf{Direction} & \textbf{Format} \\
\midrule
\multirow[t]{5}{*}{\makecell[tl]{Sensing\\(0x0001)}}
& Capabilities & 0x0001 & Read
& \makecell[tl]{\texttt{u8} flags\\bit 0: tactile present} \\

& IMU Data & 0x0002 & Notify
& \makecell[tl]{\texttt{u8} sample\_count\\per sample: \texttt{i16}$\times$3 accel, \texttt{i16}$\times$3 gyro, \texttt{u32} timestamp} \\

& Tactile Data & 0x0005 & Notify
& \makecell[tl]{\texttt{u32} timestamp\\\texttt{u8[64]} 8$\times$8 pressure matrix} \\

& Config & 0x0003 & Write
& \makecell[tl]{\texttt{u8} imu\_rate\_hz\\\texttt{u8} tactile\_rate\_hz} \\

& Echo & 0x0004 & Write/Notify
& - \\
\midrule
\multirow[t]{3}{*}{\makecell[tl]{Feedback\\(0x0002)}}
& Motor Command (Stop) & 0x0001 & Write
& \makecell[tl]{\texttt{u8} cmd = 0x00\\\texttt{u8} motor\_id} \\

& Motor Command (Play) & 0x0001 & Write
& \makecell[tl]{\texttt{u8} cmd = 0x01, \texttt{u8} motor\_id\\\texttt{u16} duration\_ms, \texttt{u8} count\\\texttt{u16} gap\_ms, \texttt{u8} intensity} \\

& Echo & 0x0002 & Write/Notify
& - \\
\bottomrule
\end{tabularx}
\vspace{-10pt}
\end{table*}
All stickies, shown in Fig.~\ref{fig:pcb-details}, are built around the nRF52832 Bluetooth Low Energy SoC. We selected the nRF52832 for its low-power operation and because its required pins remained routable within the PCB assembly constraints of the accessible manufacturer. The sensing sticky, detailed in Table~\ref{tab:sensing_components_cost}, integrates a BMI323 6-axis IMU. We use 6-axis rather than 9-axis inertial sensing to avoid magnetometer sensitivity to local magnetic disturbances and to reduce power consumption. The sensing sticky also exposes an expansion interface for the optional tactile extension. The extension integrates an external tactile sensing circuit using 74HC4051 row and column multiplexers (Texas Instruments) and supports customizable tactile patches constructed from a Velostat layer sandwiched between flexible PCB traces. The feedback sticky, detailed in Table~\ref{tab:feedback_components_cost}, integrates a DRV8836 motor driver and supports up to two VC0720B015F eccentric rotating mass (ERM) vibration motors.

Each sticky is powered by a detachable 70 mAh LiPo battery. We selected detachable LiPo batteries over smaller or more integrated alternatives because practical recharging, battery replacement, and day-to-day reuse mattered more than achieving the minimum possible thickness. The electronics are enclosed in breathable, water-resistant, and tear-resistant pouches from Scratch Free Packaging, made from non-woven spunbond polypropylene. The pouches keep the reusable electronics from directly contacting the skin while allowing the adhesive interface to be selected and replaced for each task or wearer. Across our experiments and studies, we attached the pouches using skin-safe kinesiology tape, waterproof film bandages, and waterproof silicone bandages.

\subsection{Firmware and BLE Services}
\begin{figure}[t]
\includegraphics[width=0.75\columnwidth]{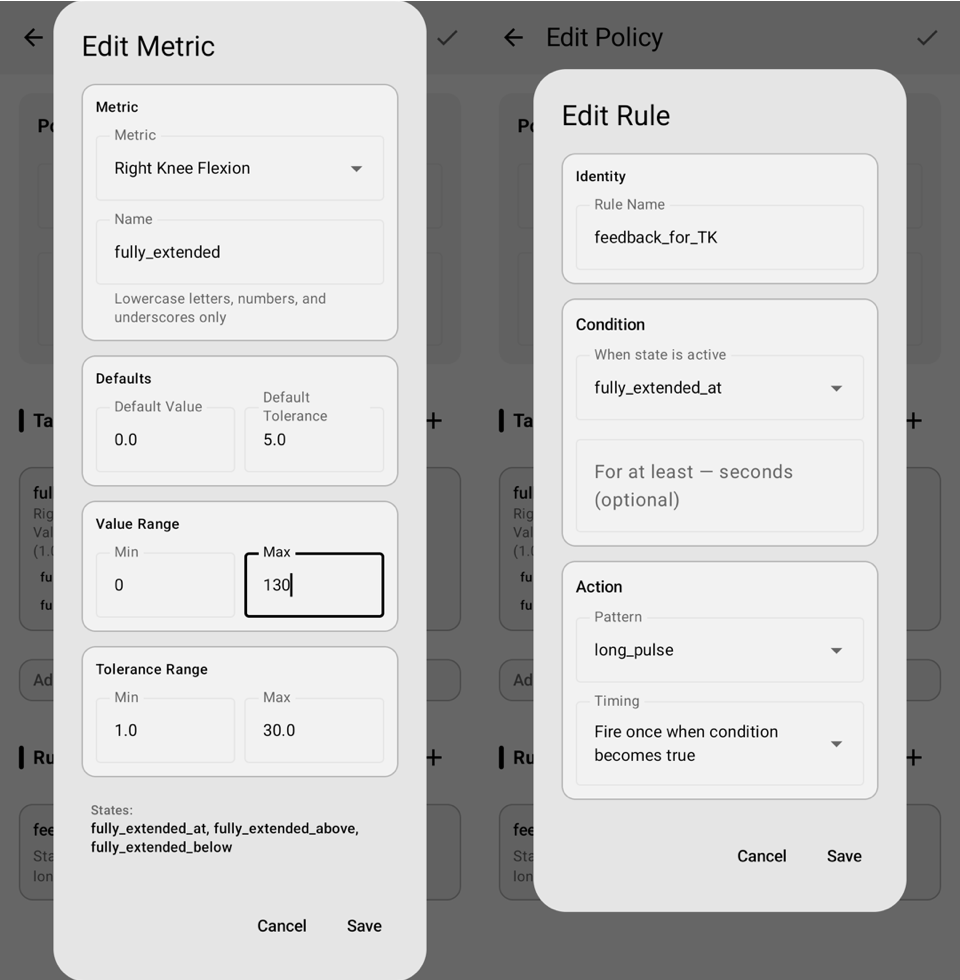}
  \caption{Screenshots of metric and rule editing screens.}
  \label{fig:edit-screenshots}
\end{figure}
The embedded implementation keeps firmware simple, reusable, and energy-conscious by moving task-specific sensing interpretation and feedback-policy logic to the companion app. The platform uses two role-specific firmwares on the same BLE platform: one for sensing stickies and one for feedback stickies. 

Both firmwares expose device identity and capabilities over BLE, allowing the app to distinguish sticky types without manual device-type selection. On boot and when a BLE connection is established, sensing firmware detects whether a tactile extension is attached by scanning a subset of tactile channels and thresholding their ADC responses.

We release prebuilt firmware binaries that only need to be flashed once after assembly. Firmware is programmed through an SWD interface using a pogo-pin probe clip and a J-Link programmer.

Table~\ref{tab:ble_gatt} summarizes the BLE GATT structure. On the sensing side, BLE services expose device capabilities, streamed IMU data, streamed tactile data when an extension is present, runtime sensing-rate configuration, and an echo path used for latency measurement. On the feedback side, BLE services expose motor commands and a corresponding echo path. 

Sensing firmware defers IMU initialization until a BLE connection is established. Sampling is timer-driven, and the IMU and tactile sampling rates can be reconfigured at runtime through BLE commands from the app. In the current implementation, IMU streaming runs at 100 Hz and tactile streaming runs at up to 30 Hz. Feedback firmware renders vibration patterns locally and places the motor driver into sleep mode between patterns to reduce idle power consumption.
\subsection{Companion App and Configuration Representation} \label{appendix-companion-app}
The companion mobile app is implemented as an Android application in Kotlin. It maps physical stickies into the body-centered configuration model, instantiates tasks, feedback policies, and session configurations in software, and provides the authoring, setup, calibration, and runtime machinery for executing them. Physical devices are assigned to anatomical locations so that users work with body- and task-level terms rather than raw device identifiers.
The three configurable objects described in Sec.~\ref{sec:platform-overview} are represented internally as follows:
\begin{itemize}[leftmargin=*, nosep]
\item \textbf{Task representations} store sensing requirements, anatomical placements, calibration flows, and the metrics made available for policy authoring and runtime evaluation. The implemented joint metrics are computed from IMUs on two adjacent body segments and include task-relevant joint angles and angular velocities, such as flexion, carrying angle, and pronation. Segment metrics are computed from one IMU and include orientation-derived quantities such as pitch and roll, as well as motion-derived quantities such as angular velocity and linear acceleration. These quantities can be exposed through body-centered concepts such as forward/backward or side-to-side leaning. Tactile metrics include contact presence, total and peak tactile values, contact area, and center-of-pressure coordinates.
\item \textbf{Feedback-policy representations} operate over these metrics rather than over raw sensor streams. A policy defines named states as boolean conditions over metric values. More complex states can be composed by referencing other states, logical relationships among states, or state durations. Conditions can reference multiple metrics simultaneously, including metrics derived from multiple joints or body segments. When a condition is satisfied, its associated action specifies a vibration pattern to be rendered by one or more feedback stickies.
\item \textbf{Session-configuration representations} instantiate a task-policy pair for a concrete use session. They store selected feedback locations, tuned parameter values, and vibration settings. Session-level changes can therefore modify both the state or condition parameters of the selected policy and the way its actions are rendered on the body, without redefining the sensing task or authoring a new policy from scratch.
\end{itemize}
At runtime, the app receives sensor streams, computes the configured metrics and states, evaluates policy conditions, and sends vibration commands to the selected feedback stickies. The current policy runtime is designed for real-time rules over current metric and state values together with short-lived state history. It does not support arbitrary time-series queries, such as the average elbow angle over the preceding 10 seconds, or general multi-phase procedural scripts, such as ``first do X, then Y,'' unless the required sequence is encoded through explicit task- or policy-specific states. The current implementation likewise does not support learned policies such as pose classification.

\begin{figure}[t]
\includegraphics[width=0.75\columnwidth]{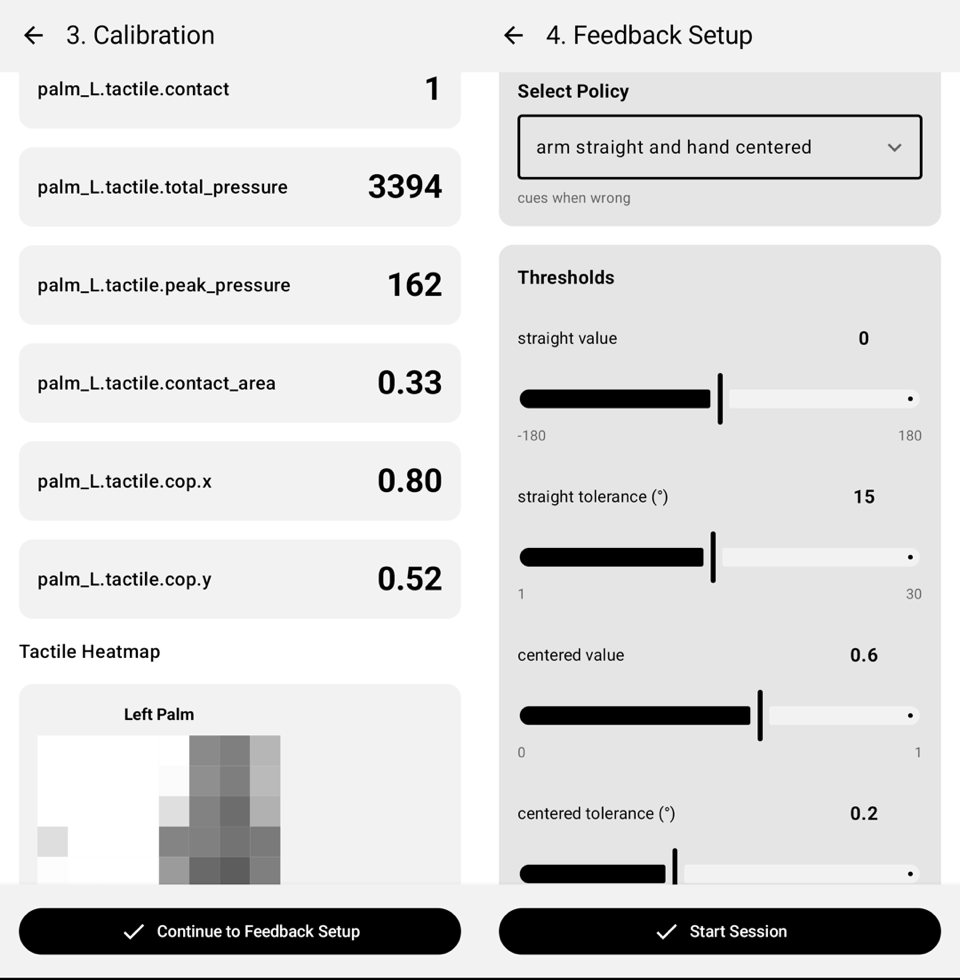}
  \caption{Screenshots of tactile patch metrics and policy used for the end-user study.}
  \label{fig:tactile-screenshots}
\end{figure}

\begin{figure}[t]
\vspace{-10pt}
\includegraphics[width=\columnwidth]{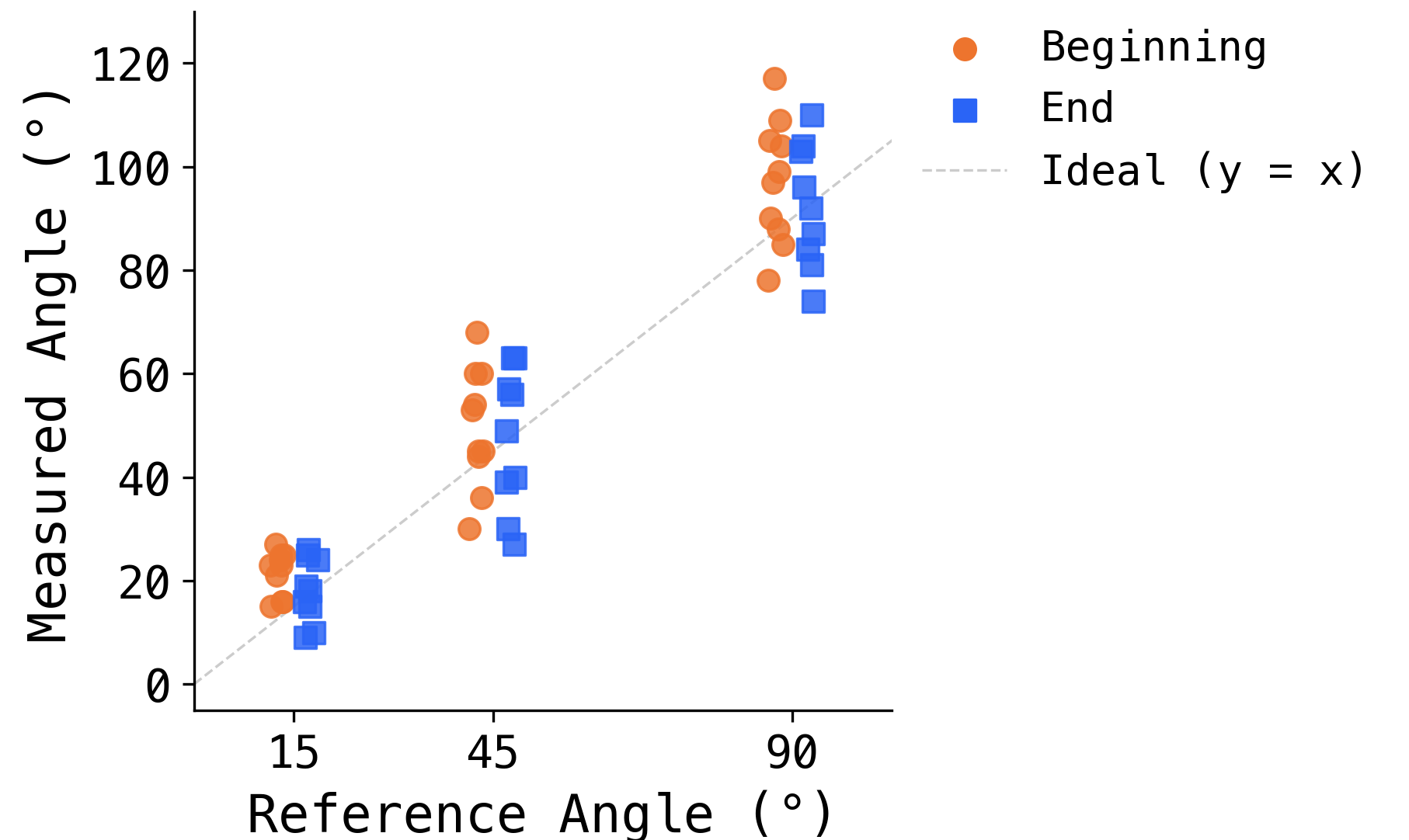}
  \caption{Left-elbow flexion feasibility check for end-user setup and calibration. App-estimated angles at 15°, 45°, and 90° goniometer references are shown for measurements taken after initial setup and at the end of the session, for each participant. The dashed line indicates ideal agreement. Results suggest that participants could place and calibrate the sensing stickies well enough to support the task, while still showing participant-specific offsets and spread.}
  \label{fig:sensing-accuracy}
\end{figure}
\begin{figure*}[t]
\includegraphics[width=0.95\textwidth]{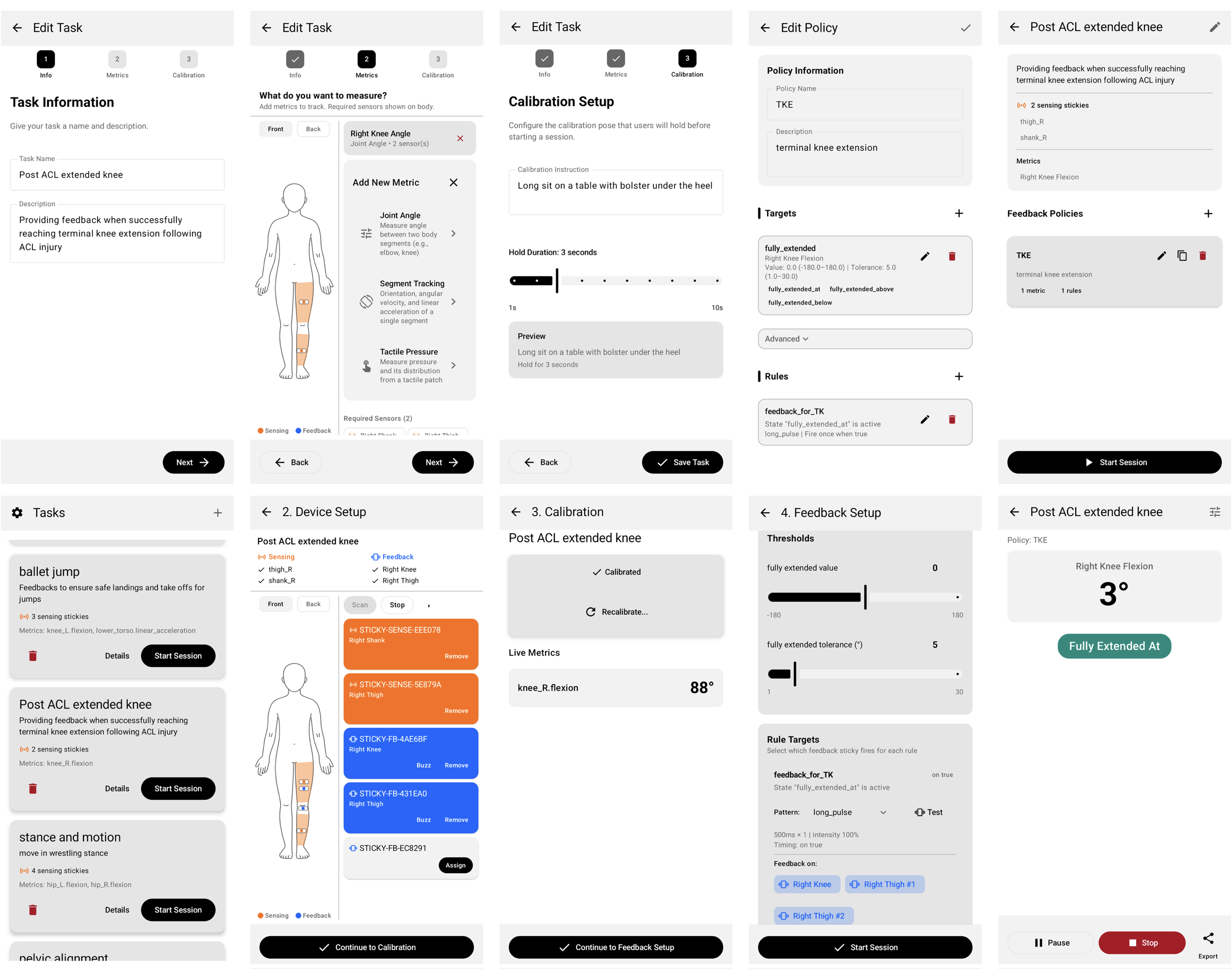}
\vspace{-10pt}
\caption{Screenshots of the task and policies co-configured with E4, the athletic trainer, during the study.}
  \label{fig:ui_screenshots}
\end{figure*}

\subsection{Sensing Pipeline and Calibration}\label{appendix-sensing-pipeline-calibration}
The body-state metrics exposed by the app are implemented through metric-specific sensing and calibration pipelines.
Orientation estimation runs independently on each sensing sticky at 100 Hz using VQF~\cite{vqf}, a quaternion-based 6-axis orientation filter with gyroscope-bias estimation and rest detection. We ported the algorithm from and checked its behavior against the open-source reference implementation (\url{https://github.com/dlaidig/vqf}).
Segment metrics are derived from calibrated body-relative orientations under practical placement assumptions. In particular, the system treats the IMU board normal as a proxy for the local skin normal. A user-specified reference pose establishes the body-relative frame from which segment-orientation metrics are computed.

Tactile processing uses baseline subtraction to compensate for resting sensor values. Patch-orientation calibration then maps the tactile matrix into the selected body-relative orientation so that raw tactile readings can be converted into contact, pressure-summary, contact-area, and center-of-pressure metrics.

Joint-metric estimation additionally compensates for unknown sensor-to-body alignment and heading drift between paired sensing stickies~\cite{1dof-algo, 2dof-algo}. During a short calibration movement, typically consisting of repeated joint bends and rotations, the system estimates the functional joint axes of the instrumented body segments. For 1-DoF joints, the joint axis is identified using constraint-based optimization~\cite{1dof-algo}. For 2-DoF joints, the system estimates two joint axes together with their heading offset~\cite{2dof-algo}. The pipeline also performs online heading correction and reference-pose body-frame calibration. Joint angles are then computed from the relative orientations of the calibrated body frames and decomposed into the task-relevant degrees of freedom exposed by the app.
The current implementation does not incorporate a complete anatomical model for 3-DoF joints. For joints such as the shoulder and hip, it supports constrained task movements by treating the relevant motion as a 1-DoF joint task.

Calibration is implemented as a multi-phase state machine. It verifies that the required sensor streams are available, captures the movements and reference poses required by the selected inertial metrics, and records tactile baselines and patch orientation when tactile sensing is used. Users can selectively recalibrate an individual joint, the reference pose, or the tactile baseline without repeating the full calibration procedure.

\section{Technical Characterization Details}\label{sec:appendix-characterization}
\subsection{BLE Latency and Packet Loss Details}
We characterized BLE performance across 17 device setups that varied the numbers of sensing, tactile-enabled sensing, and feedback stickies. Each setup was measured for two minutes per condition. Because the Android phone used in the study (Pixel 9A) supported at most 8 concurrent BLE connections, many measurements focused on configurations near that limit.

During each trial, the wearer performed walking, jumping jacks, and arm swings while the stickies were adhered to the skin and covered by a hoodie. The phone was either carried on the body, in the hand or pocket, or placed 5 m away. Sensing stickies streamed 100 Hz IMU data, and tactile-enabled sensing stickies additionally streamed 30 Hz tactile data.

Latency was estimated using periodic echo messages every 3 s and computed as the sum of sensing-to-phone and phone-to-feedback delay. This measure captures BLE communication latency rather than full closed-loop end-to-end latency: it excludes sensing-side processing, app-side decision logic, and actuator onset. However, BLE communication is a major contributor to end-to-end delay in multi-sticky deployments.

As shown in Fig.\ref{fig:characterization}(A), the smallest setup, with one sensing sticky and one feedback sticky, showed higher latency than several denser configurations. We suspect this reflects variability in how Android's BLE stack negotiates connection intervals across different numbers of concurrent connections, which can vary by device and OS version. More broadly, however, adding stickies did not produce a systematic latency increase.

\begin{figure}[t]
\vspace{-3pt}
\includegraphics[width=\columnwidth]{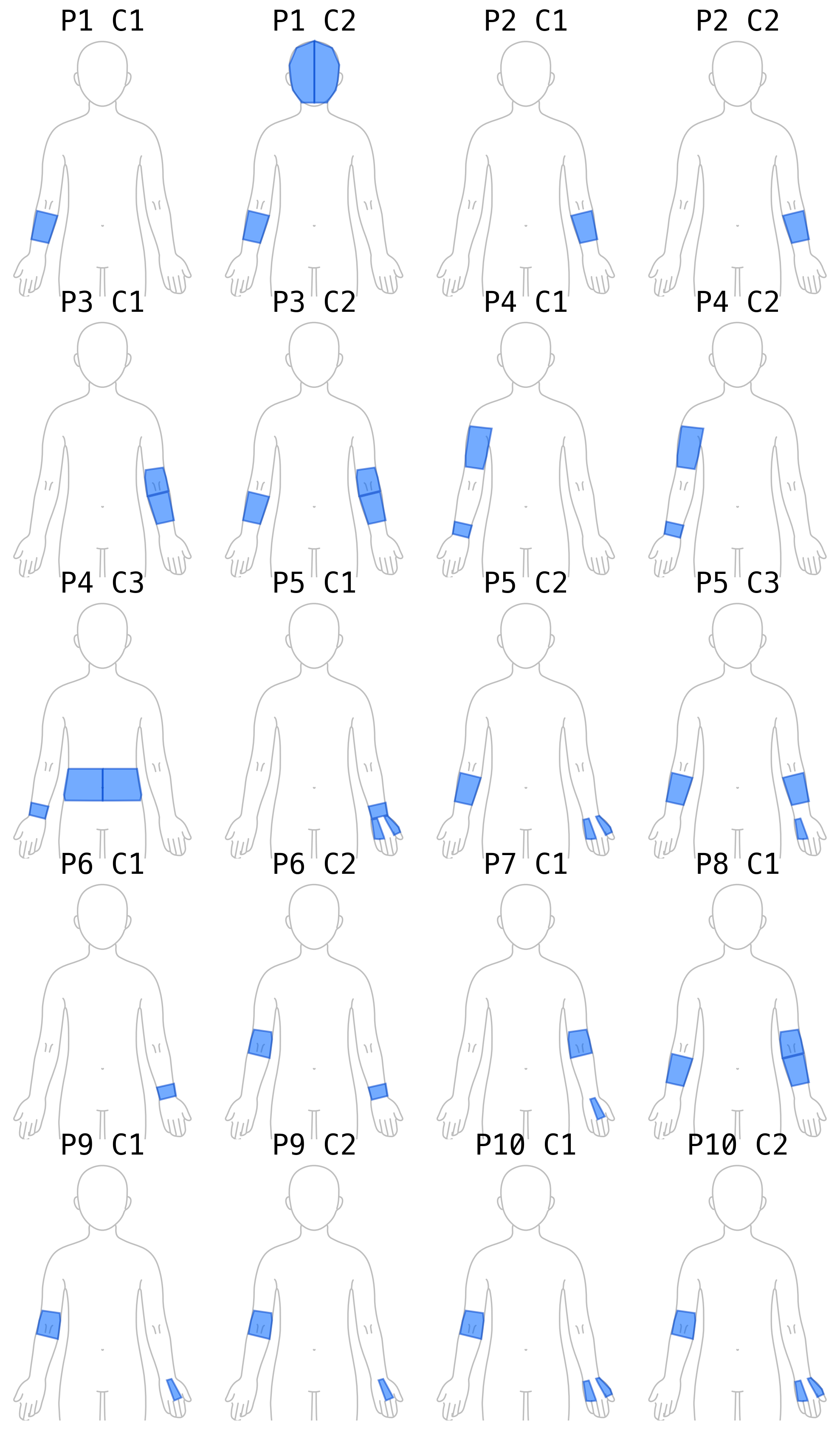}
\vspace{-25pt}
  \caption{All configurations from the end-user configuration studies. ``Px Cy'' denotes the `x'th participant's `y's configuration. Two consecutive configurations are the same when participants only tuned parameters or changed vibration settings in the session config.}
  \label{fig:study-configurations}
\end{figure}
\renewcommand{\arraystretch}{1.08}
\setlength{\tabcolsep}{3.0pt}
\newcolumntype{L}{>{\raggedright\arraybackslash}X}

\begin{table*}[h]
\caption{Example training tasks expressed through \thePlatform{}'s shared configuration model. Prior work used task-specific form factors, while \thePlatform{} realizes analogous sensing, feedback, and policy logic by reconfiguring the same sensing and feedback stickies.}
\label{tab:example-task-configs}
\centering
\small
\begin{tabularx}{\textwidth}{
  >{\raggedright\arraybackslash}p{0.12\textwidth}
  >{\raggedright\arraybackslash}p{0.24\textwidth}
  >{\raggedright\arraybackslash}p{0.20\textwidth}
  >{\raggedright\arraybackslash}p{0.20\textwidth}
  L
}
\toprule
\textbf{Task} & \textbf{Prior-work form factor(s)} & \textbf{Sense} & \textbf{Feedback} & \textbf{Feedback policy} \\
\midrule

Balance Training &
Belts with IMUs and vibration motors~\cite{vibrotactile-for-standing} &
1 sensing sticky on lower back: trunk orientation relative to neutral standing &
4 feedback stickies around waist &
If trunk pitch or roll exceeds a direction-specific threshold, pulse the spatially corresponding sticky to cue left/right/front/back sway. \\

\midrule

Arm Elevation Ergonomics &
Customized workwear T-shirt integrated with two IMUs, with one vibration actuation unit on a semi-elastic textile strap~\cite{imu-vibrotacrtile-ergonomics-intervention} &
2 sensing stickies on left and right upper arms: segment orientations for arm elevation relative to arms-hanging calibration &
1 feedback sticky on dominant arm &
If either upper-arm elevation exceeds a lower or higher limit (30\textdegree{} or 60\textdegree{}), trigger the corresponding warning level with different vibration patterns. \\

\midrule

Knee Angling for Snowboarding &
Force sensors in shoes, bend sensor on knee, IMU on arm, and vibration motors on body~\cite{Wearable-automatic-feedback-snowboard} &
2 sensing stickies on left thigh and left shank: knee flexion angle and thigh orientation relative to natural standing &
2 feedback stickies on left knee and pelvis &
Use thigh orientation as a standing gate. When standing, cue when knee flexion crosses personalized flexion or extension targets; suppress cues on lifts or after falls. \\

\midrule

CPR Compression Training &
Glove with tactile matrix on palm and back of hand, with vibration motors on wristband~\cite{CPR-glove}; strapped plate on the hand relying on audiovisual feedback~\cite{CPR-band}&
2 sensing stickies on left forearm and left upper arm, with tactile patch on back of left hand: elbow flexion angle and center of pressure's $x$-coordinate &
Variable feedback layout: users choose the number and placement of feedback stickies &
Cue when the elbow is too bent, when the top hand is too left, or when the top hand is too right, using personalized thresholds and mappings. \\

\bottomrule
\end{tabularx}
\vspace{-10pt}
\end{table*}
\subsection{Energy Consumption Details}
In advertising mode, base sensing and feedback stickies drew 80 uA on average, corresponding to 36.5 days of standby lifetime.
With the tactile patch extension, advertising current increased to 564 uA, or 5.2 days, due to the additional ADC circuitry.
Once connected and streaming, a sensing sticky transmitting 100 Hz IMU data drew 979 uA on average, corresponding to 3.0 days of continuous operation. 
Enabling the tactile patch increased average current to 3.96 mA, reducing estimated lifetime to 17.7 h (18.8 h from real-world measurements) because tactile scanning keeps the MCU active longer and larger packets increase BLE radio time, though this still supports roughly a full day of continuous use.
For feedback stickies, as shown in Fig.~\ref{fig:characterization}(B), energy consumption depended on duty cycles. 
During vibration, peak current was approximately 65 mA, and the average current during active buzzing was 45 mA. 
Because feedback is typically sparse and event-driven, long-term average current is much lower: with one 100 ms buzz per minute, the average current was 237 uA, corresponding to 12.2 days of use.
A feedback sticky could support about 11 or 50 buzzes per minute while staying within the average power profile of a connected sensing sticky, depending on if it is tactile-enabled, exceeding the needs for sparse cues during training.
\end{document}